\documentclass[aps,pra,twocolumn,notitlepage,nofootinbib,superscriptaddress]{revtex4-1}

\usepackage{rahulbhadani}
\usepackage{amsmath}
\usepackage{amssymb}
\usepackage{graphicx}
\usepackage[colorlinks=true,allcolors=blue]{hyperref}
\usepackage{braket}
\usepackage{dsfont}
\usepackage{multirow}
\usepackage{subcaption}
\usepackage{listings}
\begin{document}

\title{On Simulation of Power Systems and Microgrid Components with SystemC-AMS}

\author{Rahul Bhadani}
\email{rahul.bhadani@uah.edu}
\affiliation{The University of Alabama in Huntsville}

\thanks{This work was partly funded by the Advanced Research
Projects Agency-Energy (ARPA-E), U.S. Department of Energy, under Award Number DE-AR0001580.}

\author{Satyaki Banik}
\email{sbanik@ncsu.edu}
\affiliation{North Carolina State University}

\author{Hao Tu}
\email{htu@ncsu.edu}
\affiliation{North Carolina State University}

\author{Srdjan Lukic}
\email{smlukic@ncsu.edu}
\affiliation{North Carolina State University}

\author{Gabor Karsai}
\email{gabor.karsai@vanderbilt.edu}
\affiliation{Vanderbilt University}

\begin{abstract}
Cyber-physical systems such as microgrids consist of interconnected components, localized power systems, and distributed energy resources with clearly defined electrical boundaries. They can function independently but can also work in tandem with the main grid. Power system converters and their control loops play an essential role in stabilizing grids and interfacing a microgrid with the main grid. The optimal selection of microgrid components for installation is expensive. Simulation of microgrids provides a cost-effective solution. However, when studying the electromagnetic transient response, their simulation is slow. Furthermore, software packages facilitating electromagnetic transient response may be prohibitively expensive. This paper presents a faster method for simulating the electromagnetic transient response of microgrid components using SystemC-AMS. We present a use case of a photovoltaic grid-following inverter with a phase-locked loop to track reference active and reactive power. Our results demonstrate that the simulation performed using SystemC-AMS is roughly three times faster than the benchmark simulation conducted using Simulink. Our implementation of a photovoltaic grid-following inverter equipped with a phase-locked loop for monitoring reference active and reactive power reveals that the simulation executed using SystemC-AMS is approximately three times faster than the benchmark simulation carried out using Simulink. Our implementation adopts a model-based design and produces a library of components that can be used to construct increasingly complex grid architectures. Additionally, the C-based nature allows for the integration of external libraries for added real-time capability and optimization functionality. We also present a use case for real-time simulation using a DC microgrid with a constant resistive load.
\end{abstract}

\maketitle

\section{Introduction}
\label{sec:intro}

\begin{figure}[htpb]
\centering
\includegraphics[width=1.0\linewidth, trim={0cm 0.0cm 0cm 0.0cm},clip]{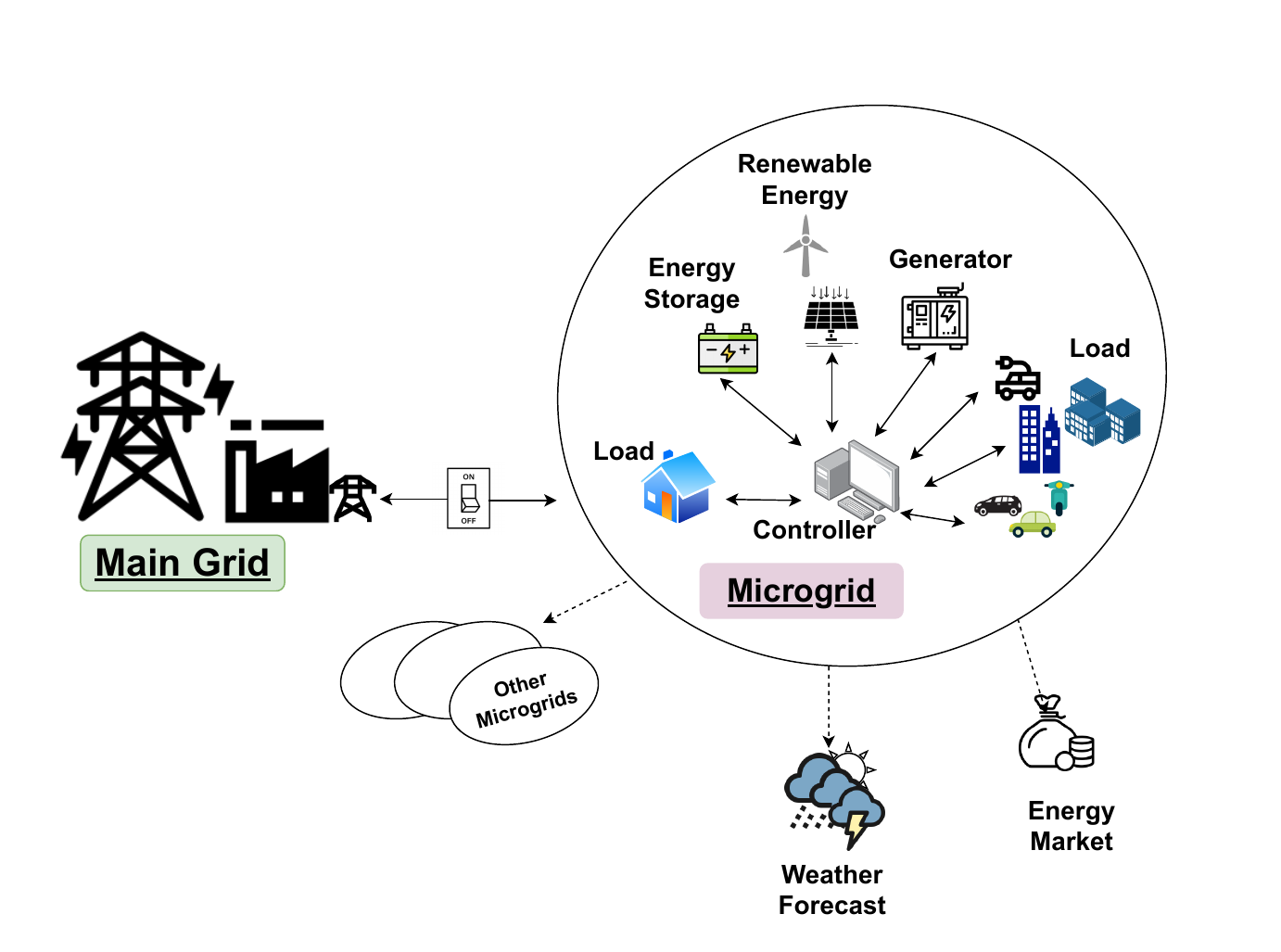}
\caption{An illustration of a microgrid with a photovoltaic renewable energy source.}
\label{fig:microgrid}
\end{figure}

Modern electric grids incorporate various computational and communication components integrated with digital controllers, making them suitable for study as Cyber-Physical Systems (CPS). In recent years, the approach to electric grid design has changed due to the rapid increase in the share of renewable energy and the proliferation of Internet of Things (IoT) components. This has led to stricter grid interconnection requirements by grid operators to improve grid flexibility and stability. Photovoltaic (PV) and wind power plants can be connected to the grid through power converters, which not only transfer generated DC power to the AC grid but also provide services such as dynamic control of active and reactive power, reactive current injection during faults, and the ability to control grid voltage and frequency~\cite{rodriguez2021grid}. These renewable energy sources along with various loads and other components form a small-scale grid called microgrids (Fig. ~\ref{fig:microgrid}). A microgrid is designed to serve a small community such as a university campus or an office space. It can connect to and disconnect from the grid to operate in two forms -- grid-connected mode, and islanded mode~\cite{booth2020microgrids, farrokhabadi2019microgrid}. Microgrids can improve the main grid's resiliency due to grid disturbance and improve customer reliability.

\begin{figure}[htpb]
\centering
\includegraphics[width=1.0\linewidth, trim={0cm 0.0cm 0cm 0.0cm},clip]{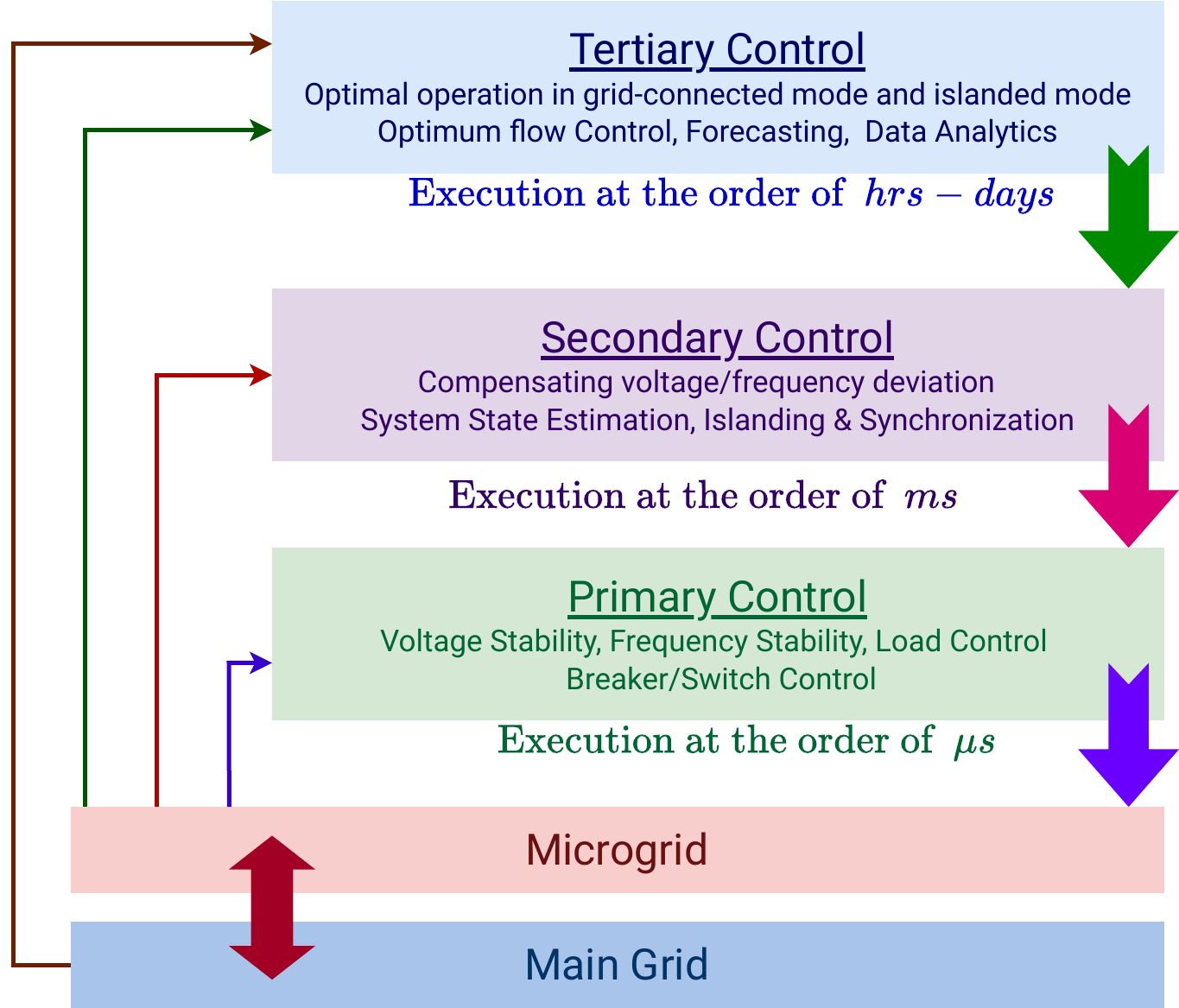}
\caption{Hierarchical control in a microgrid, operating at varying degrees of time-scale.}
\label{fig:Controls_In_Microgrids}
\end{figure}

As a microgrid is a complex system, directly installing such a system without studying its feasibility via computer simulation might be costly and may not yield optimal results. Hence, a digital twin or a simulation tool to study the microgrid's performance with various component selections is highly sought after. Over the years, several methods for simulating microgrid components have been proposed~\cite{alzahrani2017modeling}. These methods range from equation-based models to neural network-based models and provide simulation capabilities at varying degrees of granularity, from electromagnetic transients (EMT)\cite{xu2014fast, nzimako2016real} to economic interests\cite{che2014dc, sandelic2022reliability}. This aligns with the hierarchical control of a microgrid, which has three levels:
\begin{enumerate}
    \item \textbf{Primary Control:} focuses on the immediate and local management of microgrid components, ensuring stable voltage and frequency by responding to real-time changes in load and generation. It involves decentralized actions like droop control and inertia emulation to maintain system stability. 
    \item \textbf{Secondary Control:} provides more refined adjustments to restore any deviations caused by primary control, ensuring the system returns to its nominal operating conditions. This level involves coordinated actions, such as automatic generation control (AGC) and voltage regulation, typically managed by a centralized controller.
    \item \textbf{Tertiary Control:} oversees the economic and optimal operation of the microgrid, addressing the broader aspects of energy management, such as power flow optimization, market participation, and coordination with the main grid. This level integrates advanced algorithms for energy scheduling, demand response, and resource optimization to enhance the overall efficiency and cost-effectiveness of the microgrid.
\end{enumerate}
Microgrid control operations are illustrated in Figure~\ref{fig:Controls_In_Microgrids}.

While studying simulation at the level of economic interests is limited to developing optimization routines and mathematical modeling, EMT simulation requires detailed modeling of microgrid components. In this paper, we focus on simulating the EMT of a microgrid using SystemC-AMS (AMS stands for Analog/Mixed-Signal)~\cite{IEEESTD, vachoux2004towards, barnasconi2010systemc}, which is shown to be roughly three times faster than Simulink-based simulation. Such an improvement in simulation time is desirable, as the evaluation and analysis of EMT simulations may take long hours to accomplish for large grid systems with time-varying signals.

The main contribution of this paper is the use of SystemC-AMS for the simulation of power systems and microgrids that exhibit electromagnetic transients. We demonstrate the use of SystemC-AMS for microgrid simulation using a detailed model of a grid-following inverter for PV. We provide two variations of grid-following inverters: one using a low-pass filter and one without a low-pass filter. We adopt the model-based design for creating simulations in SystemC-AMS, where we first separately develop components in SystemC-AMS that can be used to construct a microgrid along with controllers to achieve a desired objective. Additionally, we provide a use case of a real-time simulation based on SystemC-AMS that can facilitate hardware-in-the-loop simulation.

\section{Background and Literature Review}
The concept of microgrids is an elegant way to integrate renewable energy sources with the main grid using power converters and inverters. However, the use of power converters and other electrical components adds electromagnetic transients that need to be studied in both islanded and grid-connected modes. Converter-based generators have faster switching properties and require faster control compared to traditional synchronous generators. Thus, the control and protection of microgrids~\cite{xu2014fast} remain a challenging problem. Microgrid designers and power engineers employ computer simulation to study the effect of EMT.

Several simulation tools have been proposed in the past to study EMT. OPAL-RT, a real-time simulator developed by a Canadian company, provides the capability to perform EMT simulation~\cite{bian2015real}. Several researchers have also proposed the use of RTDS simulators for power systems and microgrids~\cite{xu2014fast, kuffel1995rtds, logenthiran2012multiagent}. However, both OPAL-RT and RTDS require specialized hardware for conducting EMT simulation. There are also non-real-time simulation tools available for conducting EMT simulations. Among them, the most popular is Simulink with the `Specialized Power Systems' toolbox~\cite{sybille2000digital}. Simulink provides a specialized library for creating microgrids in the simulation. A detailed analysis of power system components using Simulink requires greater computational cost and time. Another recent method called DPSIM for EMT simulation is based on dynamic phasor~\cite{mirz2019dpsim}. Dynamic phasor-based simulation allows conducting EMT simulation in the phasor domain, contrary to other time-domain-based simulations. In such a simulation, results need to be converted back to the time domain after executing the simulation. As of writing this report, the use of DPSIM is limited and doesn't allow the use of user-defined controllers and components to create arbitrarily complex microgrids.

Various other software toolchains exist that simulate power systems and microgrids at different levels. OpenDSS is an open-source simulation package for simulating power systems~\cite{montenegro2012real}. It is capable of performing steady-state simulation and quasi-steady-state simulation. OpenDSS may not be suitable for dynamic simulation, including EMT simulation.

In this paper, we use SystemC-AMS C++ libraries for microgrid simulation. SystemC-AMS is based on IEEE standard 1666.1-2016 and provides several models of computation (MoC) including Timed Data Flow (TDF) and Electrical Linear Network (ELN). We use TDF and ELN MoC to create several grid components and controllers such as three-phase voltage sources, transmission lines, phase-locked loops, and low-pass filters. SystemC-AMS is a mixed-analog extension of SystemC, originally designed as a library for the design and verification of hardware systems~\cite{panda2001systemc}. Using SystemC-AMS, the discrete event simulation of SystemC can be extended to support continuous time systems~\cite{vachoux2004towards}. Furthermore, SystemC-AMS clearly differentiates between conservative and non-conservative models. Such differentiation helps model electrical circuits that obey Kirchhoff's circuit laws.

The remaining part of the paper is organized as follows. In Section~\ref{sec:systemcams}, we describe how grid components can be modeled using SystemC-AMS. In Section~\ref{sec:gfl}, we describe the PV Grid-following  (PV-GFL) Inverter and how it can be used to stabilize active and reactive power. Section~\ref{sec:results} provides a comparison of the execution time of SystemC-AMS-based simulation with the execution time obtained from Simulink. Section~\ref{sec:real_time} provides a use-case on real-time simulation using SystemC-AMS and ZeroMQ library.
Finally, we provide a conclusion and the future outlook.

\section{Modeling Grid Components using SystemC-AMS}
\label{sec:systemcams}
\begin{figure}[htpb]
\centering
\includegraphics[width=1.0\linewidth, trim={0cm 0.0cm 0cm 0.0cm},clip]{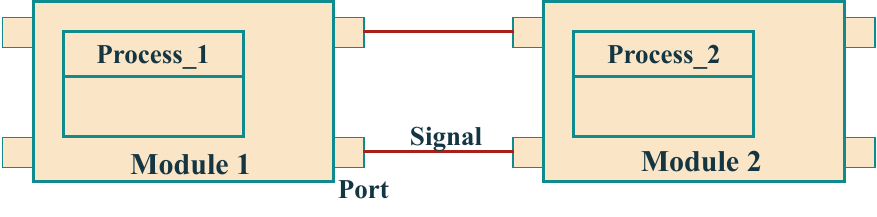}
\caption{An illustration showing modules, ports, signals, and processes in SystemC. }
\label{fig:SystemC_schematic}
\end{figure}

SystemC was developed as a standardized library to enable system-level design and sharing of semiconductor intellectual property (IP) core. SystemC provides modeling construct similar to hardware description languages such as VHDL and Verilog~\cite{swan2001introduction}. SystemC allows the construction of structural designs using modules, ports, and signals. All ports and signals are declared to be of specific data types. They also enable data communication between modules. Modules, through processes, can implement any desired functionality. They are designed to be concurrent. Core C++ doesn't provide a notion of time or clock, however, SystemC has a built-in notion of a Clock that can be used to create discrete event simulation. Fig.~\ref{fig:SystemC_schematic} shows how modules, ports, and signals are related to each other in SystemC.

SystemC-AMS is an extension of SystemC designed to simulate mixed and analog systems such as electrical circuits and continuous domain transfer functions. It is capable of simulating both discrete-time and continuous-time systems. Additionally, it can distinguish between conservative behaviors (such as the conservation of energy and Kirchhoff's laws) and non-conservative behaviors (such as high-level abstraction and event-driven modeling). Physical quantities like voltage or current are associated with conservative behavior as they follow physical laws such as Kirchhoff's voltage and current laws. A Model of Computation (MoC) defines a set of rules for behavior and interaction between SystemC-AMS primitive modules. SystemC-AMS supports three kinds of MoC: (1) Timed-Data Flow (TDF); (2) Electrical Linear Networks (ELN); and (3) Linear Signal Flow (LSF).

TDF MoC can be used to model discrete event systems and corresponding simulations without using the expensive dynamic scheduling imposed by SystemC's discrete event kernel. Connected TDF modules define a static schedule, forming a TDF cluster. The static schedule defines the sequence of execution, configures how many samples to read from or write to an input or output port, and specifies delays at ports. Port delays are useful for breaking algebraic loops in feedback systems.

LSF MoC is used to model non-conservative systems that are continuous in time through primitive modules such as addition, multiplication, delays, integration, etc. LSF uses differential-algebraic equations (DAE) for the implementation of primitive modules. A system of equations in the LSF model is solved using a linear DAE solver -- a part of core SystemC-AMS implementation.

ELN MoC is used to model conservative systems, continuous in time such as voltage and current where the goal is to conserve laws of physics. The value of continuous-time variables is determined in accordance with Kirchhoff's current and voltage laws using algebraic equations and is solved at run-time during simulation. SystemC-AMS only specifies a set of predefined primitives for constructing electrical circuit networks.  An ELN module can be instantiated as a child module of a SystemC parent module with the \texttt{SC\_ MODULE} macro.

In the next few subsections, we look at some components modeled using SystemC-AMS MoCs that are required for constructing a microgrid.

\subsection{Component Models for Microgrids}
\label{sec:components}
\begin{figure*}[htpb]
\centering
    \includegraphics[width=1.0\linewidth, trim={0cm 0.0cm 0cm 0.0cm},clip]{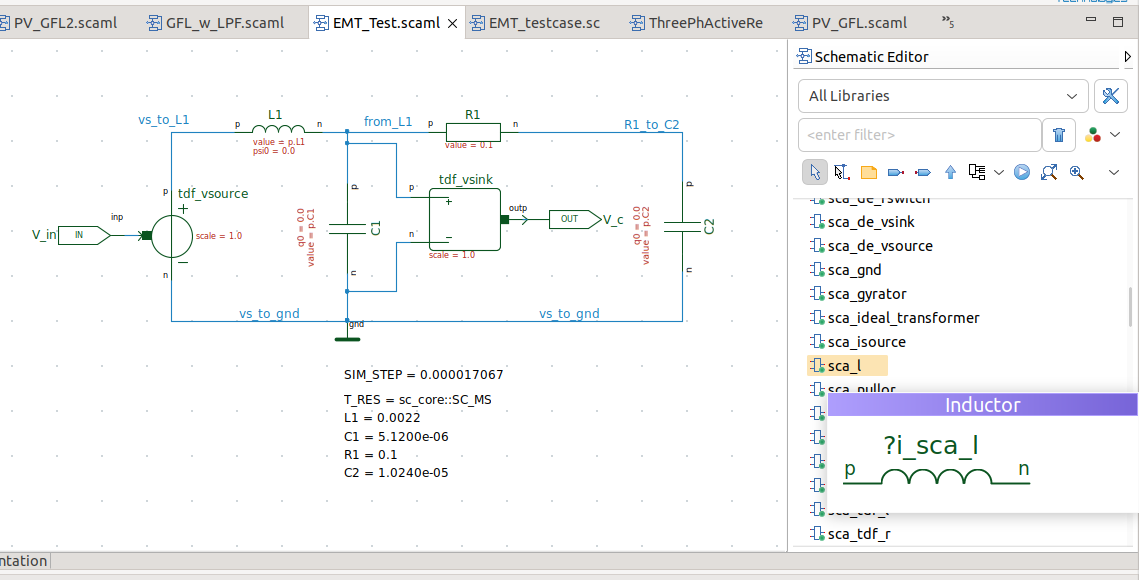}
\caption{A schematic editor in COSIDE.}
\label{fig:COSIDE_Snap}
\end{figure*}

To create microgrid components in SystemC-AMS, we use a model-based design tool called COSIDE~\cite{coside}. COSIDE provides drag-and-drop support for primitives and existing user-defined modules that generate SystemC and SystemC-AMS C++ code skeletons. Users can modify the generated code and provide their own logic in the \texttt{processing} function in the case of TDF modules. ELN primitives can be dragged and dropped on a schematic editor to construct a library block which can be reused in other schematic editors for creating hierarchically complex blocks (see Figure~\ref{fig:COSIDE_Snap} for an example). 

\subsubsection{Three-phase Voltage Source}
SystemC-AMS provides an external signal-driven voltage source primitive module. The voltage source can be used to create a three-phase voltage source by using a TDF module output as an external signal.

\begin{figure}[htpb]
\centering
\includegraphics[width=1.0\linewidth, trim={0cm 0.0cm 0cm 0.0cm},clip]{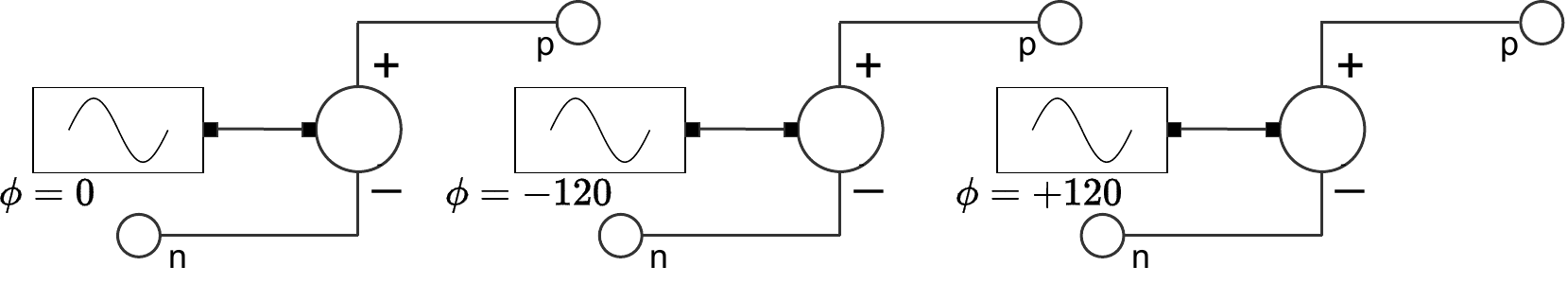}
\caption{A schematic of a three-phase voltage source.}
\label{fig:threephasev}
\end{figure}

The TDF voltage source is specified by two ELN terminals: p, for positive terminals and n for negative terminals, and a TDF input port. To create a three-phase voltage a sine-wave generator TDF block is connected to the input port of the TDF voltage source. The sine-wave generator is parametrized by amplitude, frequency, phase, sampling time, and offset. We create three such TDF voltage sources at 120-degree phases from each other. A schematic of a three-phase voltage source is shown in Fig.~\ref{fig:threephasev}.
\begin{figure}[htpb]
\centering
\includegraphics[width=1.0\linewidth, trim={0cm 0.0cm 0cm 0.0cm},clip]{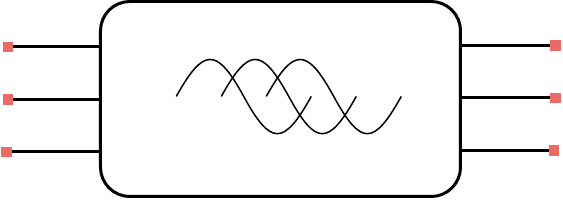}
\caption{An abstract block of three-phase voltage source from Fig.~\ref{fig:threephasev}.}
\label{fig:threephase_abs}
\end{figure}

COSIDE generates a reusable SystemC module that can be abstracted with three positive terminals and three negative terminals as abstracted in Fig.~\ref{fig:threephase_abs}. COSIDE lets the user decorate the abstracted block for reusability and an intuitive look and feel. COSIDE-generated C++ declaration of the SystemC module corresponding to the three-phase voltage source is provided in Listing~\ref{code:threephasevsource}.

\begin{lstlisting}[caption=COSIDE-generated C++ declaration of the SystemC module corresponding to the three-phase voltage source, label={code:threephasevsource}, language=C++]
SC_MODULE(ThreePhVSource)
{
/// ports
sca_eln::sca_terminal A_n;
sca_eln::sca_terminal A_p;
sca_eln::sca_terminal B_n;
sca_eln::sca_terminal B_p;
sca_eln::sca_terminal C_n;
sca_eln::sca_terminal C_p;

/// parameters
struct params
{
double RMS; /** phase-to-phase RMS voltage  */
double Frequency; /** Frequency in Hz */
double Phase; /** phase angle of phase A */
double SIM_STEP; /** Simulation Step */
unsigned int T_RES; /** Time Resolution */

params()
{
    RMS = 1000.0;
    Frequency = 100.0;
    Phase = 0.0;
    SIM_STEP = 100.0;
    T_RES = sc_core::SC_MS;
}
} p;

void architecture();
ThreePhVSource(sc_core::sc_module_name, 
	const params& pa = params());
virtual ~ThreePhVSource();
struct components;
components* c;
components* operator -> () { return c; }
};
\end{lstlisting}.

\subsubsection{Three-phase Transmission Lines}
\label{sec:transmisisonline}
\begin{figure}[htpb]
\centering
\includegraphics[width=1.0\linewidth, trim={0cm 0.0cm 0cm 0.0cm},clip]{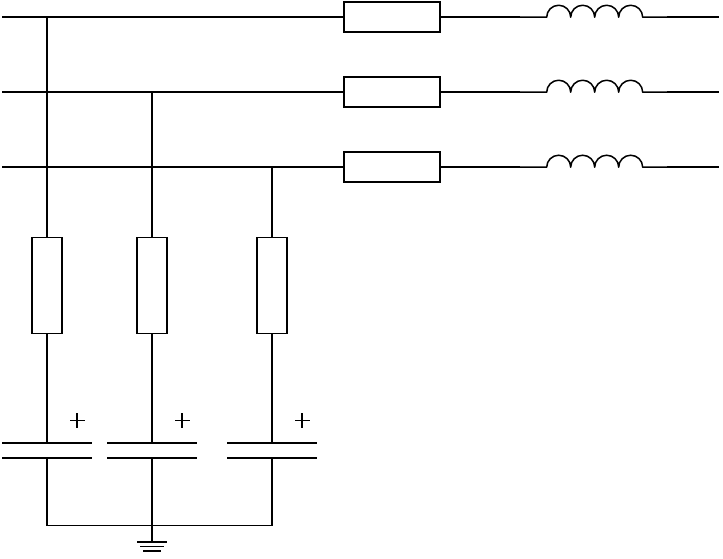}
\caption{A circuit model of loss transmission lines with series inductance and resistance acting as line impedance. We also add shunt resistance and capacitance.}
\label{fig:TranmissionLines}
\end{figure}

We model three-phase transmission lines as lossy transmission lines using resistance in series with inductance that acts as transmission line impedance~\cite{chew2020lectures}. Similarly, we have shunt capacitance to the ground along with shunt resistance in parallel. The overall circuit model of the three-phase transmission line is shown in Fig.~\ref{fig:TranmissionLines}.
Transmission lines can be modeled using pure ELN MoC. Using the COSIDE tool, we can drag and drop ELN primitives -- resistors, capacitors, and inductors to create a three-phase transmission line, which automatically generates the required SystemC-AMS code.

\subsubsection{Measuring Current and Voltage}
\begin{figure}[htpb]
\centering
\includegraphics[width=1.0\linewidth, trim={0cm 0.0cm 0cm 0.0cm},clip]{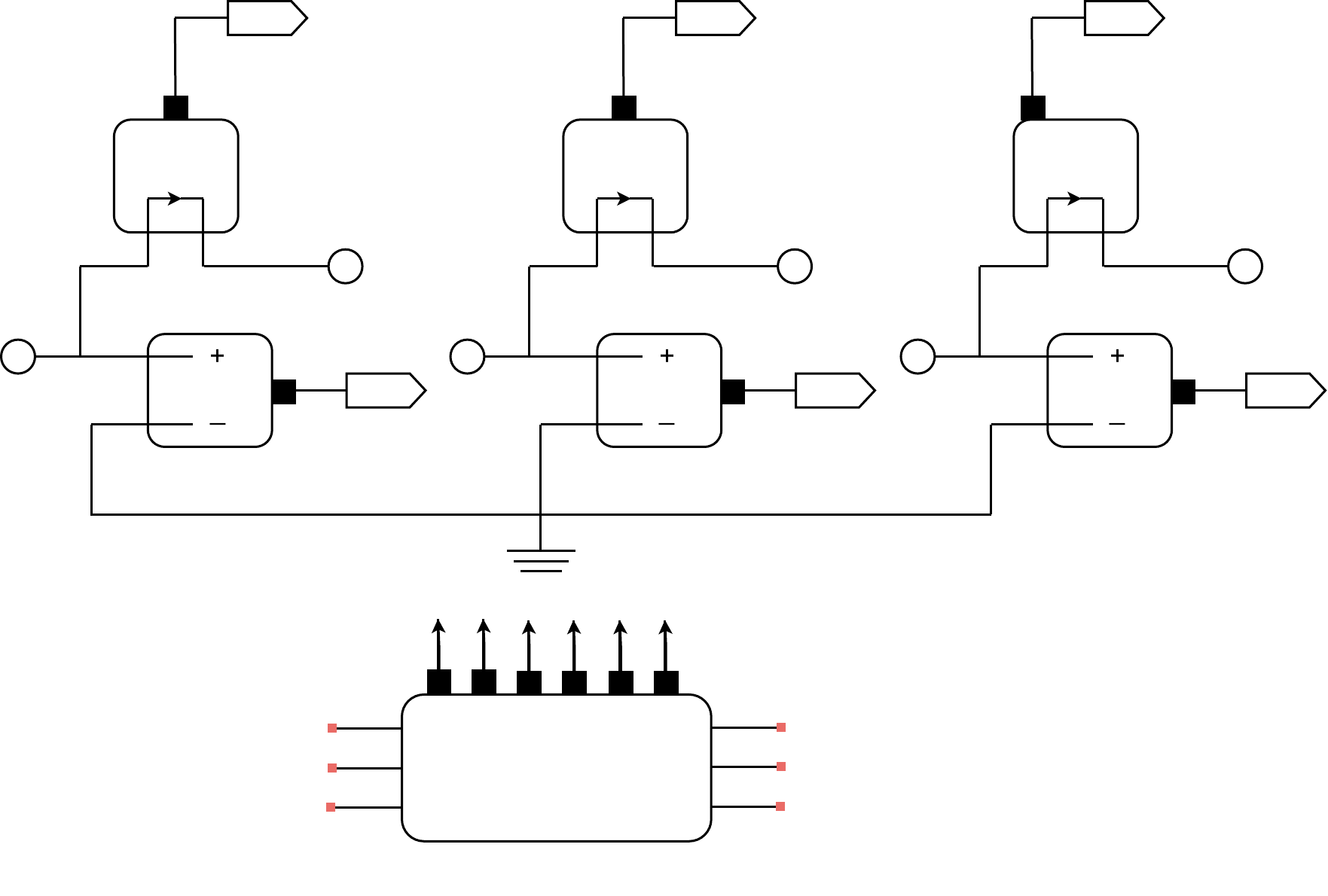}
\caption{A schematic of current and voltage measurement for three-phase electrical lines. An abstract view of the three-phase measurement block is also displayed. solid square represents a TDF port while a red square represents an electrical terminal.}
\label{fig:VIMeasurement}
\end{figure}

SystemC-AMS provides two kinds of ELN primitives for measuring the current through any electrical branch and voltage across a pair of terminals. They are called current sink and voltage sink respectively. Measuring current and voltage is required for implementing a controller for the purpose of regulating voltage and current in the grid. The schematic of a three-phase voltage and current measurement along with their abstraction is shown in Fig.~\ref{fig:VIMeasurement}.

\subsubsection{Discrete-domain Transfer Function}
To implement a digital controller, we implement a discrete-domain (or z-domain) transfer function using a TDF block. The TDF block is parameterized for re-usability to specify coefficients of the numerator as well as the denominator during the design time. Further, a user can also specify the sample time of the transfer function. A z-domain transfer function is implemented as a discrete difference equation shown in Eq.~\eqref{eq:difference_eq} in the \texttt{processing} function of a TDF module.
\begin{equation}
\begin{split}
    y[n] = -\sum_{k = 1}^N a_k y[n-k] + \sum_{k = 1}^M b_k x[n-k]
    \end{split}
    \label{eq:difference_eq}
\end{equation}
where $N$ is the number of coefficients in the denominator of the z-domain transfer function, and $M$ is the number of coefficients in the numerator of the z-domain transfer function.

\begin{figure}[htpb]
\centering
\includegraphics[width=1.0\linewidth, trim={0cm 0.0cm 0cm 0.0cm},clip]{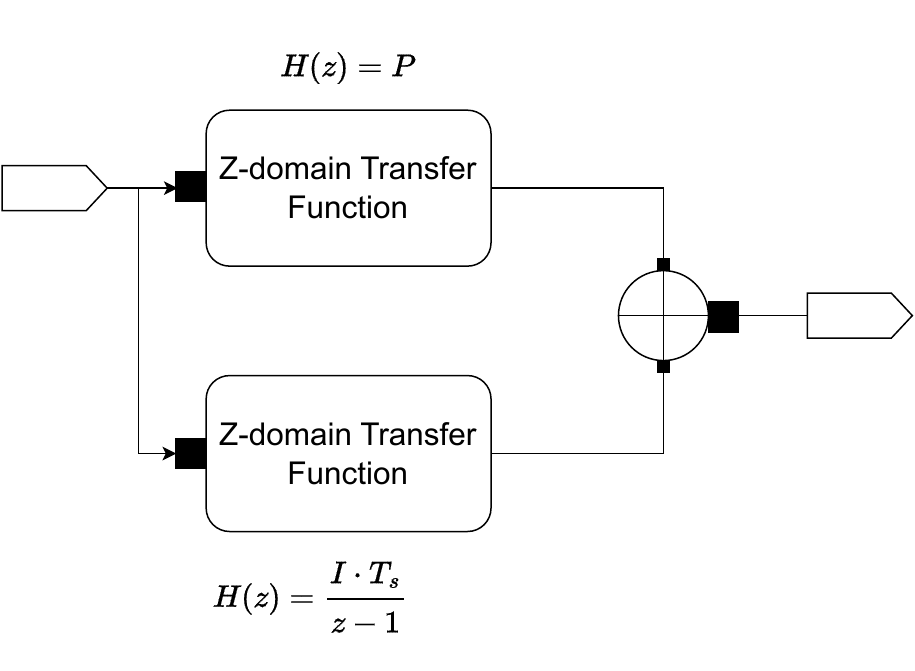}
\caption{A schematic of z-domain parallel PI controller using reusable z-domain TDF module.}
\label{fig:PI}
\end{figure}

We reuse the z-domain TDF module to create a PI controller module implemented as a parallel discrete PI controller as shown in Fig.~\ref{fig:PI}. Similarly, an arbitrary z-domain transfer function-based module such as a low-pass filter can be implemented using the z-domain TDF module.

\subsubsection{Continuous Transfer Function or s-domain Transfer Function}
SystemC-AMS provides a TDF primitive class \texttt{sca\_ltf\_nd} to implement the s-domain Laplace transfer function. We can further modularize the use of \texttt{sca\_ltf\_nd} in a TDF module's \texttt{processing} function to implement a continuous transfer function with coefficients of numerator and denominator specified by a user at the design time. Additionally, we can implement mathematical operations such as integration in the form of $\frac{1}{s}$ using \texttt{sca\_ltf\_nd}.

\subsubsection{abc to dq0 Reference Frame Converter}
A GFL inverter controls the AC side of the current and follows the phase angle of the grid voltage through a phase-locked loop (PLL). PLL implementation is described in Section~\ref{sec:pll}.  As AC signals may have time-varying phasors, converting them to a constant quantity is helpful for analysis. We use the park transformation to convert a time-varying three-phase signal, represented by abc to dq0 (direct-quadrature-zero). dq0 transformed signals lead to simpler dynamical models and are easier to analyze. A system of equations shown in Eq.~\eqref{eq:abc2dq0} is used in the \texttt{processing} function of a TDF module to implement an $abc-dq0$ converter.

\begin{equation}
\begin{split}
    d & = \frac{2}{3} \bigg(  a \sin( \omega(t) )  + b \sin( \omega(t) - 2\frac{\pi}{3 } ) + c \sin( \omega(t) + 2\frac{\pi}{3 } ) \bigg) \\
    q & = \frac{2}{3} \bigg(  a \cos( \omega(t) )  + b \cos( \omega(t) - 2\frac{\pi}{3 } ) + c \cos( \omega(t) + 2\frac{\pi}{3 } ) \bigg)\\
    z & = \frac{1}{3} \bigg(a + b  + c\bigg)
    \end{split}
    \label{eq:abc2dq0}
\end{equation}
where $\omega$ is estimated by PLL. abc signals are fed to the TDF module through the input port. To convert back $dq0$ quantities to the rotating reference frame, we implement Equation~\eqref{eq:dq02abc} in the \texttt{processing} function of a TDF module.
\begin{equation}
\begin{split}
    a & = d \sin( \omega(t)) + q \cos( \omega(t)) + z \\
b & = d \sin( \omega(t) - 2\frac{\pi}{3 }) + q \cos( \omega(t) - 2\frac{\pi}{3 } ) + z \\
c & = d \sin( \omega(t) + 2\frac{\pi}{3 }) + q \cos( \omega(t) + 2\frac{\pi}{3 }) + z \\ 
    \end{split}
    \label{eq:dq02abc}
\end{equation}

\subsubsection{Phase-locked Loop}
\label{sec:pll}
PLL measures the voltage phase angle by controlling the q-component of the three-phase voltage, after converting to $dq0$ reference frame, to zero through a PI controller. PLL establishes a relationship between frequency and grid voltage. It measures the voltage phase angle by controlling the $q$-component to zero through a PI controller. A GFL inverter uses PLL to keep the inverter synchronized to the main grid. The measured phase angle is then used to control the current.

In addition, we also use controlled-current source elements provided by the SystemC-AMS library that are used to simulate current sources.
\begin{figure}[htpb]
\centering
\includegraphics[width=1.0\linewidth, trim={0cm 0.0cm 0cm 0.0cm},clip]{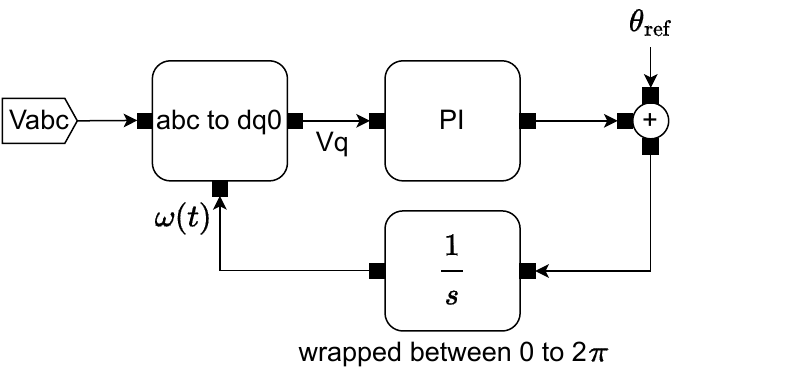}
\caption{A schematic of a phase-locked loop (PLL) used for tracking phase angle in a GFL inverter.}
\label{fig:pll}
\end{figure}
A schematic of PLL is shown in Fig.~\ref{fig:pll}. The reference from the PLL translates the three-phase instantaneous voltage obtained from the $abc$ reference frame into the rotating $dq$ reference using Park transformation. Referring to Fig.~\ref{fig:pll}, the estimated phase angle $\omega(t)$ is fed back to \texttt{abc2dq0} TDF module to drive the $q$-component of the three-phase voltage to zero. We note that when the 3-phase becomes unbalanced. the performance of the PLL design presented in Fig.~\ref{fig:pll} degrades.

\section{PV-based Grid-following Inverter for Microgrid}
\label{sec:gfl}
In terms of controller design, inverter controls can be categorized into two fundamental types: grid-following and grid-forming \cite{du2021GFLGFM}. Grid-following (GFL) control is extensively employed in grid-connected inverters, where it enables the inverter to behave akin to a current source. GFL inverters have emerged as a prominent approach for the seamless integration of distributed renewable energy into the main grid. Primarily, GFL aims to synchronize and track the grid frequency while functioning as a controlled current source operating at a designated power output. A GFL is designed to deliver the desired value of active and reactive power to the main grid. They exhibit the capability to sustain nearly constant output currents or output power during load disturbances. This active and reactive power regulation is achieved by tracking grid voltage, implementing a PLL and current control loop, allowing for rapid control of output current from the GFL \cite{IravaniBook}.

For system studies, a comprehensive photovoltaic GFL inverter model, encompassing solar cell operation and switching phenomena, is deemed to have excessive detail~\cite{li2021modeling, ge2015model}. The prevailing technique for implementing a linear controller in a three-phase system involves a  PI controller working within a dq-synchronous reference frame, wherein two independent control loops are responsible for regulating the direct and quadrature components. This study considers GFL inverters as photovoltaic (PV) units, representing the inverter side with a controllable three-phase current source cascaded with a parasitic resistance and a high-value snubber capacitance, which would absorb and dissipate high-frequency oscillations, reduce overshooting, and improve the overall transient response of the inverter \cite{xue2022siemens,du2021GFLGFM}.




\subsection{GFL control in $dq$ reference frame}
\label{sec:pv_gfl_ctrl}
For control in the $dq$ reference frame, it is important to align the $d$-axis to the space phasor of the plant model \cite{IravaniBook}. As GFL needs to lock into the frequency or phase of the grid itself, the PLL tries to achieve that through a feedback implementation which forces the $q$-axis component of inverter output voltage, $V_{sq}$ to zero. Therefore, the $d$-axis component of inverter output voltage, $V_{sd}$, becomes equal to the RMS output voltage, $\hat{V}$.
\begin{equation}
    V_{sd} = \hat{V}, V_{sq} = 0
\end{equation}

The control objective of the GFL is to regulate the real power, $P_s$, and the reactive power, $Q_s$, that is to be injected into the grid from the inverter.
\begin{equation}
    \begin{split}
        P_s(t) & = \frac{3}{2}*[V_{sd}(t) i_d(t) + V_{sq}(t) i_q(t)]\\
        Q_s(t) & = \frac{3}{2}*[-V_{sd}(t) i_q(t) + V_{sq}(t) i_d(t)]
    \end{split}
    \label{eq:power_eq1}
\end{equation}

Realizing that $V_{sq} = 0$,
\begin{equation}
    \begin{split}
        P_s(t) & = \frac{3}{2} V_{sd}(t) i_d(t)\\
        Q_s(t) & = -\frac{3}{2} V_{sd}(t) i_q(t)
    \end{split}
    \label{eq:power_eq2}
\end{equation}

From which the current references in the $dq$ domain are obtained separately,
\begin{equation}
    \begin{split}
        i_{d,ref}(t) = \frac{2}{3 V_{sd}} P_{s,ref}(t) \\
        i_{q,ref}(t) = -\frac{2}{3 V_{sd}} Q_{s,ref}(t)
    \end{split}
    \label{eq:Iref}
\end{equation}

Several of the components described in Section~\ref{sec:components} are utilized to realize two variants of the GFL inverter implementation: (1) a feedforward model, and (2) without the inner current loop and with a low-pass filter. We describe each of them below.

\subsection{Simplified GFL inverter without inner current loops}
\label{sec:pv_gfl_lpf}
\begin{figure*}[htpb]
\centering
\includegraphics[width=1.0\linewidth, trim={0cm 0.0cm 0cm 0.0cm},clip]{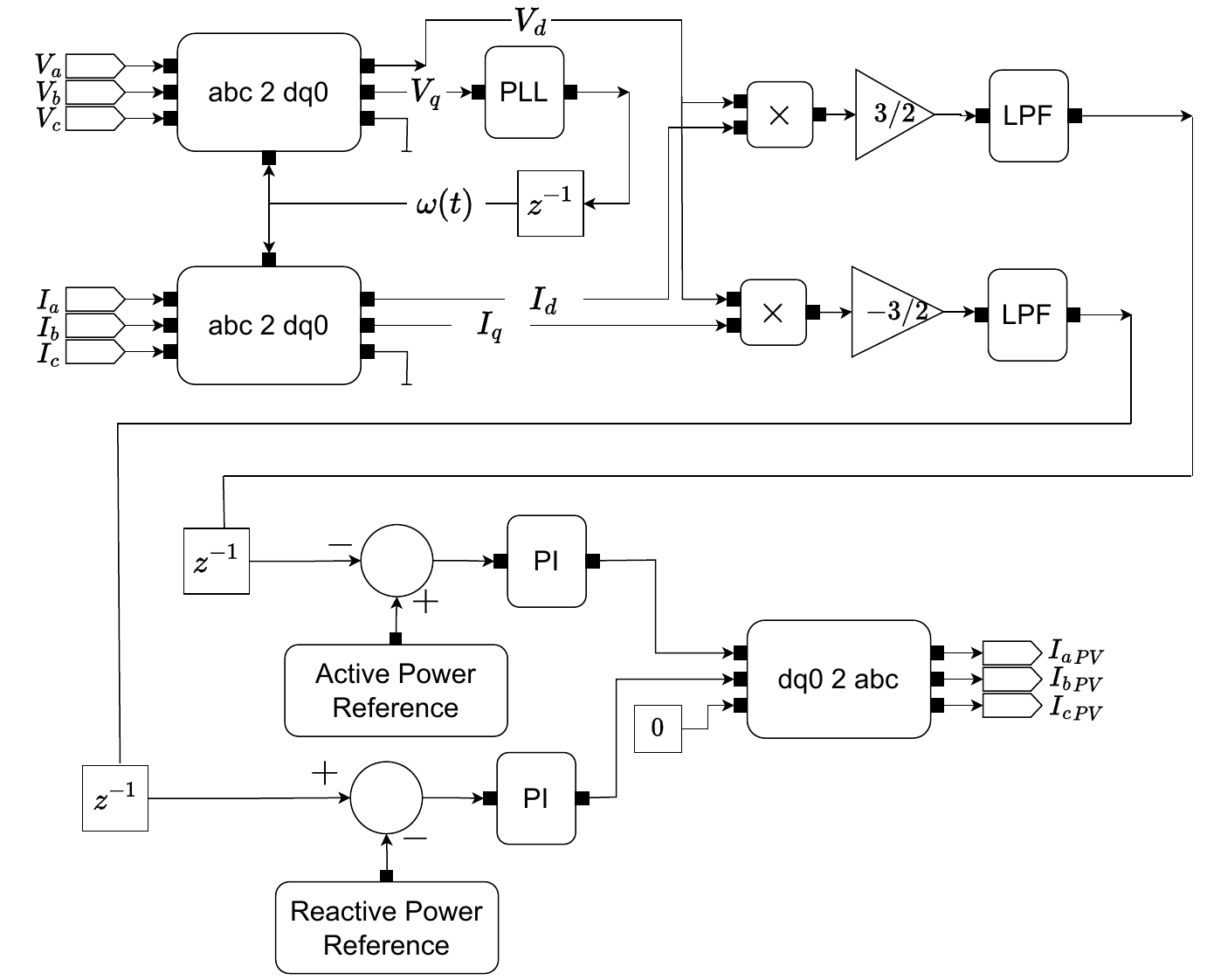}
\caption{Simplified GFL inverter without inner current loops. To break the algebraic loop that arises due to feedback, we use a delay unit $z^{-1}$ which introduces the port delay by one sample at the respective output port. }
\label{fig:PVGFL_LPF}
\end{figure*}

Without the inner current loop, the control scheme primarily consists of the outer power loop, which generates the current references for the controllable current sources. A block diagram of the simplified GFL inverter without inner current loops is provided in Fig.~\ref{fig:PVGFL_LPF}. Active and reactive power, $P$ and $Q$ are measured in the $dq$ reference frame using Eq.~\eqref{eq:power_eq1} and then filtered by a discrete low-pass filter to mitigate high-frequency components in power measurement.  The $d$ and $q$ axes are decoupled and therefore allow separate control for $P$ and $Q$. Measured power is then subtracted from the references, $P_{ref}$ and $Q_{ref}$, and then fed through a PI controller that tracks the error of active and reactive power from the specified reference to stabilize the instantaneous active and reactive power. 

\subsection{Simplified GFL inverter feedforward model}
\label{sec:pv_gfl_feed}
\begin{figure*}[htpb]
\centering
\includegraphics[width=1.0\linewidth, trim={0cm 0.0cm 0cm 0.0cm},clip]{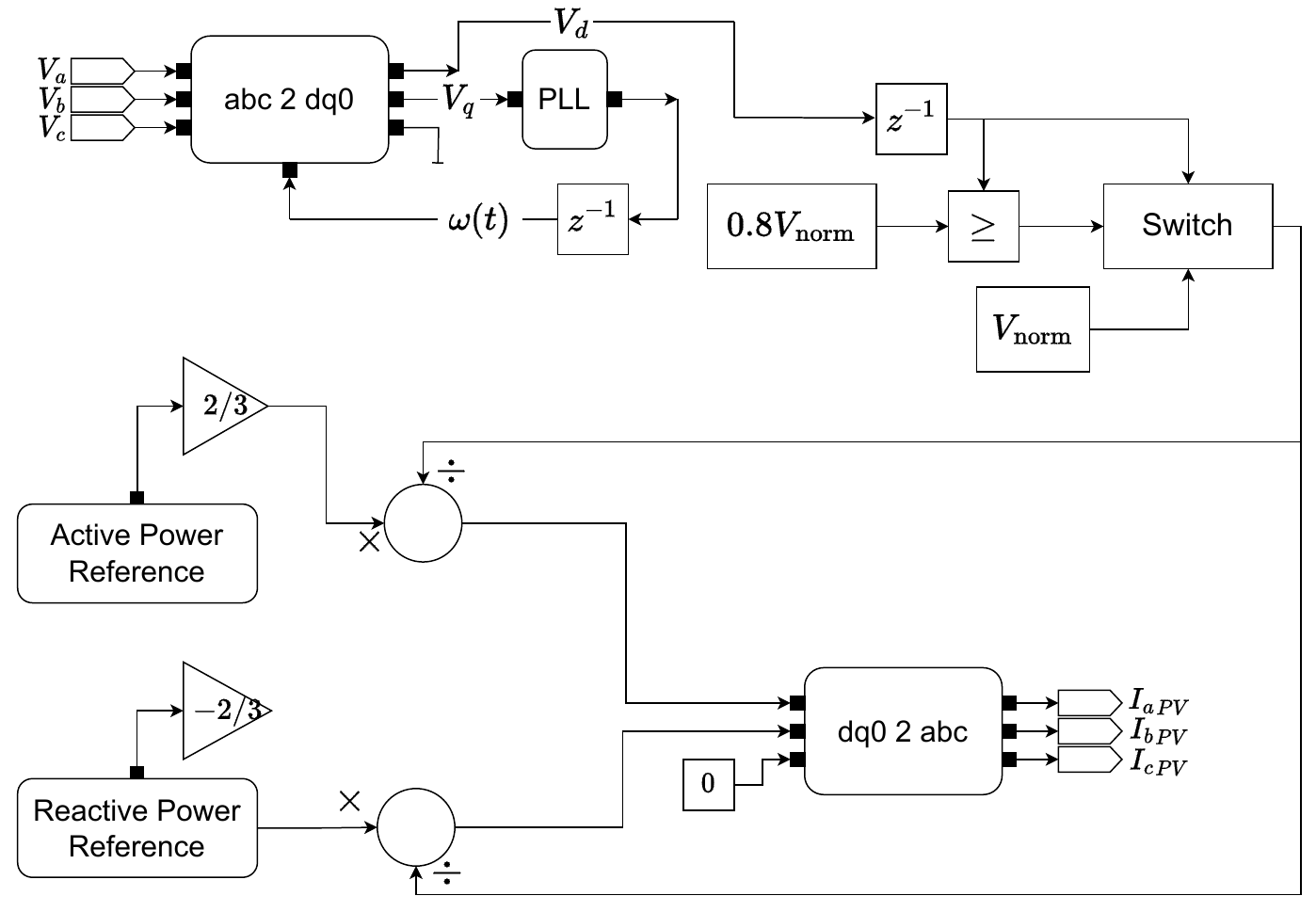}
\caption{Simplified GFL inverter feedforward model. To break the algebraic loop that arises due to feedback, we use a delay unit $z^{-1}$ which introduces the port delay by one sample at the respective output port.}
\label{fig:PVGFL_FeedFwd}
\end{figure*}

The feedforward model of the GFL inverter utilizes Eq.~\eqref{eq:Iref} to calculate the control current references directly from the power references, which is then fed into a $dq-abc$ converter block developed in Section~\ref{sec:components}. Current references in the $abc$ domain are then used as inputs to the three-phase controllable current source to inject current, and therefore power, into the grid. A block diagram of the simplified GFL inverter feedforward model is shown in Fig.~\ref{fig:PVGFL_FeedFwd}.

\begin{figure*}[htpb]
\centering
\includegraphics[width=1.0\linewidth, trim={0cm 0.0cm 0cm 0.0cm},clip]{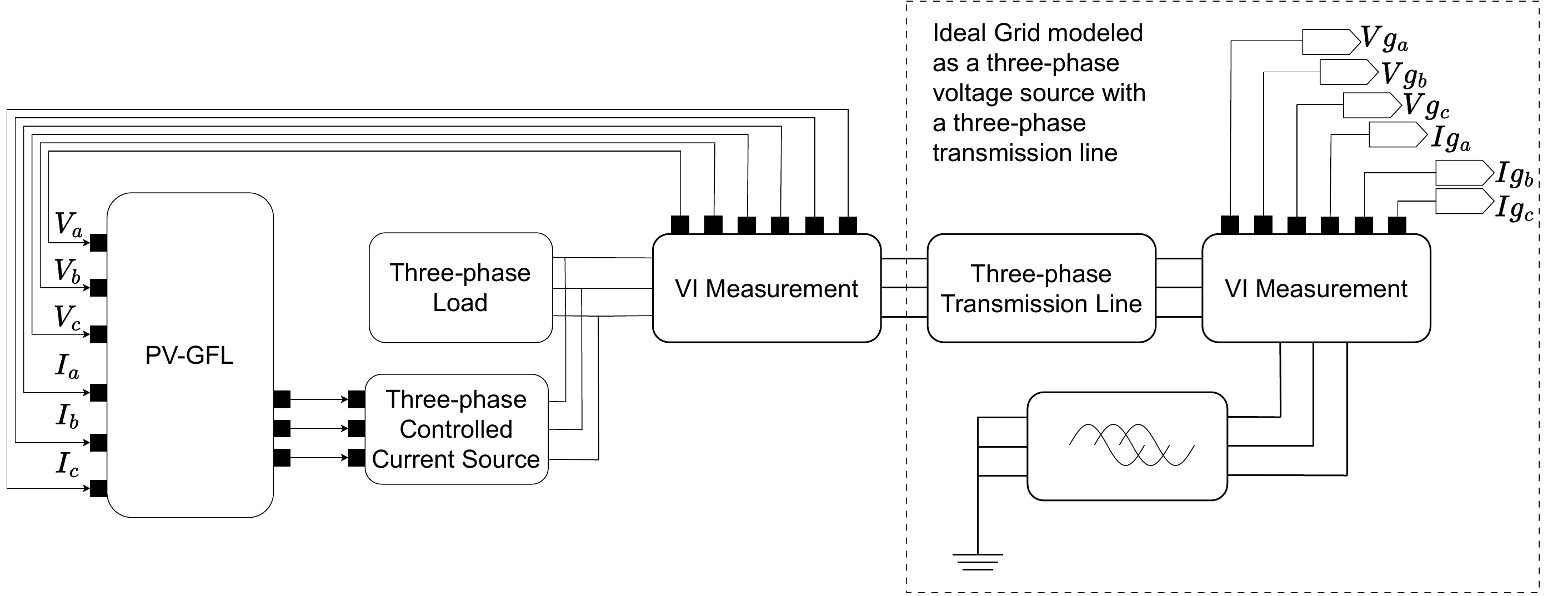}
\caption{A PV-GFL inverter acting as a microgrid that works with the main grid. The main grid is modeled as a three-phase voltage source connected through the transmission line. Three-phase controlled current source is a group of three controlled current source elements provided by the SystemC-AMS library that act as a current source based on the input value.}
\label{fig:Microgrid_architecture}
\end{figure*}

\subsection{Microgrid Architecture using PV-GFL inverter}
We can use one of the two PV-GFL designs described in Section~\ref{sec:pv_gfl_lpf} and Section~\ref{sec:pv_gfl_feed} for constructing an overall microgrid and its interaction with the main grid. The overall design is depicted in Fig.~\ref{fig:Microgrid_architecture}. We model an ideal grid as a three-phase voltage source connected through a transmission line. The transmission is a lossy transmission using resistance in series with inductance as described in Section~\ref{sec:transmisisonline}.

In the next section, we look at the simulation results and provide power measurement plots, voltage, and current plots at the inverter and grid. We also provide the execution time for comparison with Simulink.

\begin{widetext}
\begin{table}[h]
\centering
\begin{tabular}{|c|c|c|}
\hline
\textbf{Parameter} & \textbf{SystemC-AMS Simulation} & \textbf{Simulink Simulation} \\
\hline
\textbf{Simplified GFL Inverter without Inner Current Loops} & & \\
\hline
Active Power RMS Error (kW) & 85.78328 & 85.87311 \\
\hline
Reactive Power RMS Error (kVar) & 4.35805 & 4.36262 \\
\hline
\textbf{Feedforward Model} & & \\
\hline
Active Power RMS Error (kW) & 80.39742 & 80.20730 \\
\hline
Reactive Power RMS Error (kVar) & 5.92436 & 5.8789 \\
\hline
\end{tabular}
\caption{Comparison of RMS Errors for SystemC-AMS and Simulink Simulations for Step Input}
\label{tab:step}
\end{table}
\end{widetext}

\section{Results: Simulation Performance Against Simulink}
\label{sec:results}
We conducted a simulation study in SystemC-AMS for a duration of $10~\textrm{s}$ with the time-step of $50~\mu\textrm{s}$. To compare the simulation performance of microgrid implementation in SystemC-AMS, we compare the power measurement plots and several intermediate signals against Simulink implementation. For the Simulink simulation, we chose the time-step of $50~\mu\textrm{s}$ and solver settings of `fixed-step auto' which defaulted to the `ODE3' solver. In contrast, SystemC-AMS uses a linear ODE solver.

In the simulated PV-GFL, we have two kinds of reference active and reactive power that the controller needs to track: (1) step function-based reference signals; and (2) ramp function-based reference signals. As a ramp function provides a slowly varying signal, it is closer to the realistic power output from a PV. 

The three-phase voltage source chosen for the ideal grid has a root-mean-square phase-to-phase voltage of 480 V and a frequency of 60 Hz.  The transmission line is modeled with the series resistance of $0.01~\Omega$, the series inductance of $0.0001~\textrm{H}$, the shunt resistance of $0.15~\Omega$, and the shunt capacitance of $80~\mu\textrm{F}$ for each phase. The load is a pure resistive load of $1000~\Omega$ in each phase.

The low-pass filter used in the Simplified GFL inverter without inner current loops uses the z-domain transfer function given in Equation~\eqref{eq:lpf} sampling at $1000~\textrm{Hz}$.
\spliteq
{\label{eq:lpf}
\cfrac{ 0.0609}{z - 0.9391}
} 
PI Controller used for PLL described in Section~\ref{sec:pll} has a proportional coefficient of 1.088698 and an integrator coefficient of 837.46.

We first look at the step function-based reference power for two designs proposed in Section~\ref{sec:pv_gfl_lpf} and Section~\ref{sec:pv_gfl_feed}.

\subsection{Step-function Reference Power}
\label{sec:step_fn}
We study the step-response by providing a step change in the reference active and reactive power. Reference active power changes from  $0~\textrm{kW}$ to $1000~\textrm{kW}$ at $1~\textrm{s}$, $2000~\textrm{kW}$ at $2~\textrm{s}$, $1000~\textrm{kW}$ at $5~\textrm{s}$, and drops to $0~\textrm{kW}$ at $7~\textrm{s}$. Reference reactive power changes from $0~\textrm{kW}$ to $100~\textrm{kW}$ at $1~\textrm{s}$ and stays at $100~\textrm{kW}$ until the simulation ends at $t = 10~\textrm{s}$. We observe the EMT in the reactive power output as a result of the change in the active power.
\begin{figure}[htpb]
\centering
\begin{subfigure}{0.8\linewidth}
\centering
\includegraphics[width=1.0\linewidth, trim={0.0cm 0.0cm 0.0cm 0.0cm},clip]{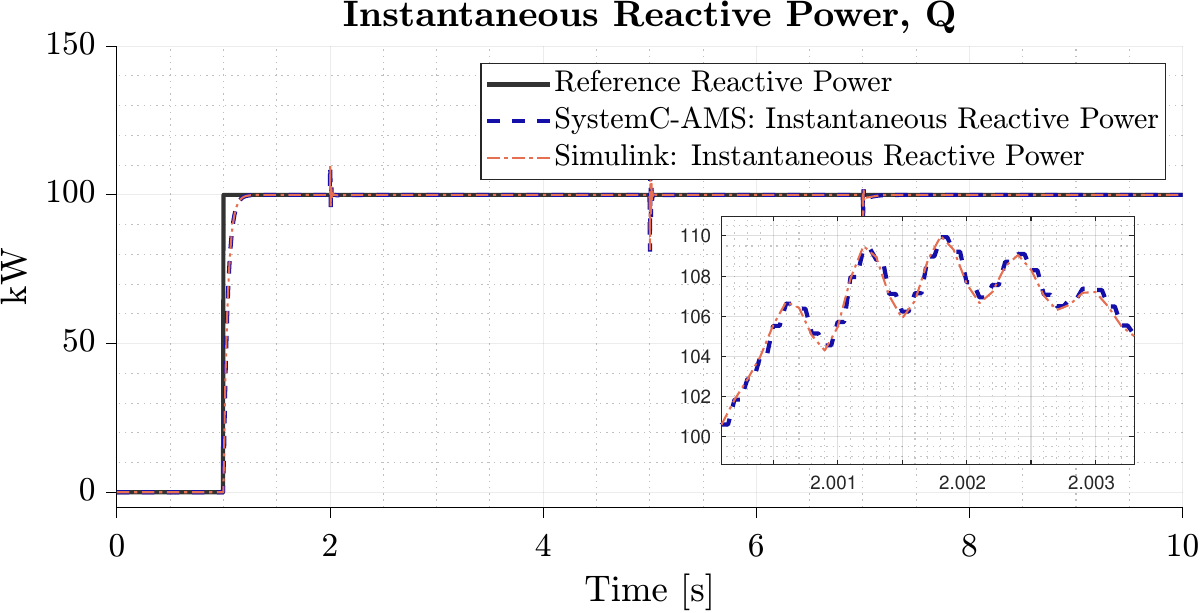}
\caption{}
\end{subfigure}
\begin{subfigure}{0.8\linewidth}
\centering
\includegraphics[width=1.0\linewidth, trim={0.0cm 0.0cm 0.0cm 0.0cm},clip]{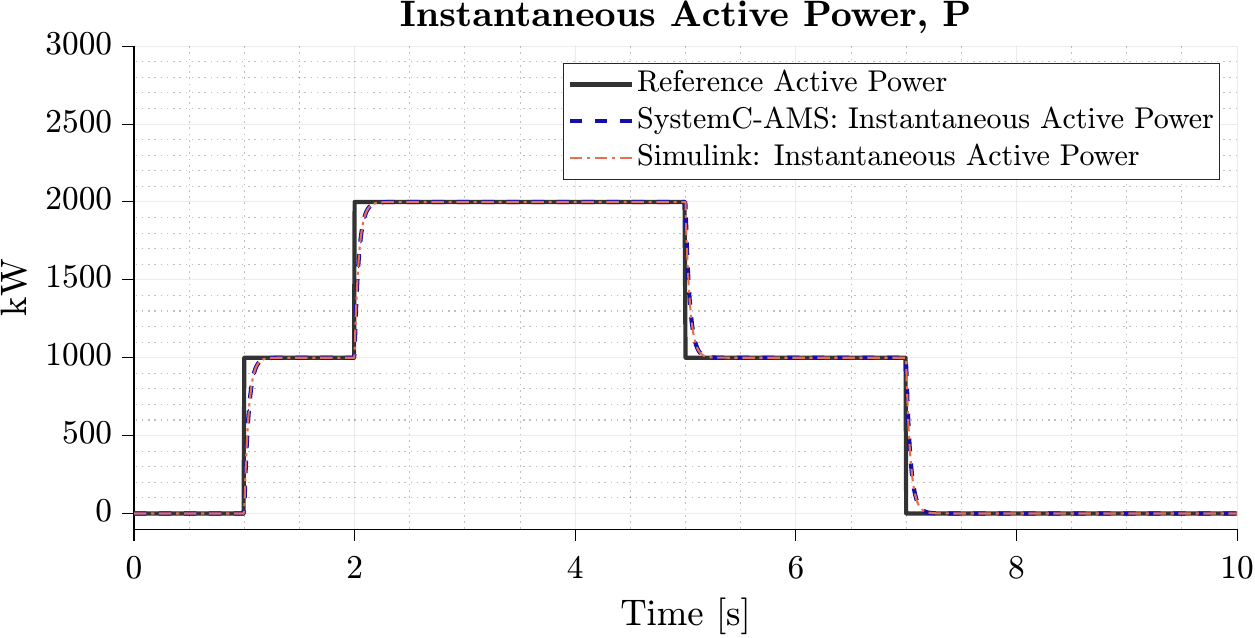}
\caption{}
\end{subfigure}
\caption{\textbf{\textit{Simplified GFL inverter without inner current loops: step response.}} \textbf{Top: }Instantaneous reactive power plot comparison in SystemC-AMS simulation vs. Simulink Simulation. We also plot the reference reactive power. \textbf{Bottom: }Instantaneous active power plot in SystemC-AMS simulation vs. Simulink Simulation. We observe the electromagnetic transients in reactive power as there is a step change in the active power -- thereby demonstrating the capability to simulate EMT.}
\label{fig:power_plot_lpf}
\end{figure}
\begin{figure}[htpb]
\captionsetup[subfigure]{aboveskip=0pt,belowskip=1pt}
\centering
\begin{subfigure}{0.8\linewidth}
\centering
\includegraphics[width=1.0\linewidth, trim={0.0cm 0.0cm 0.0cm 0.0cm},clip]{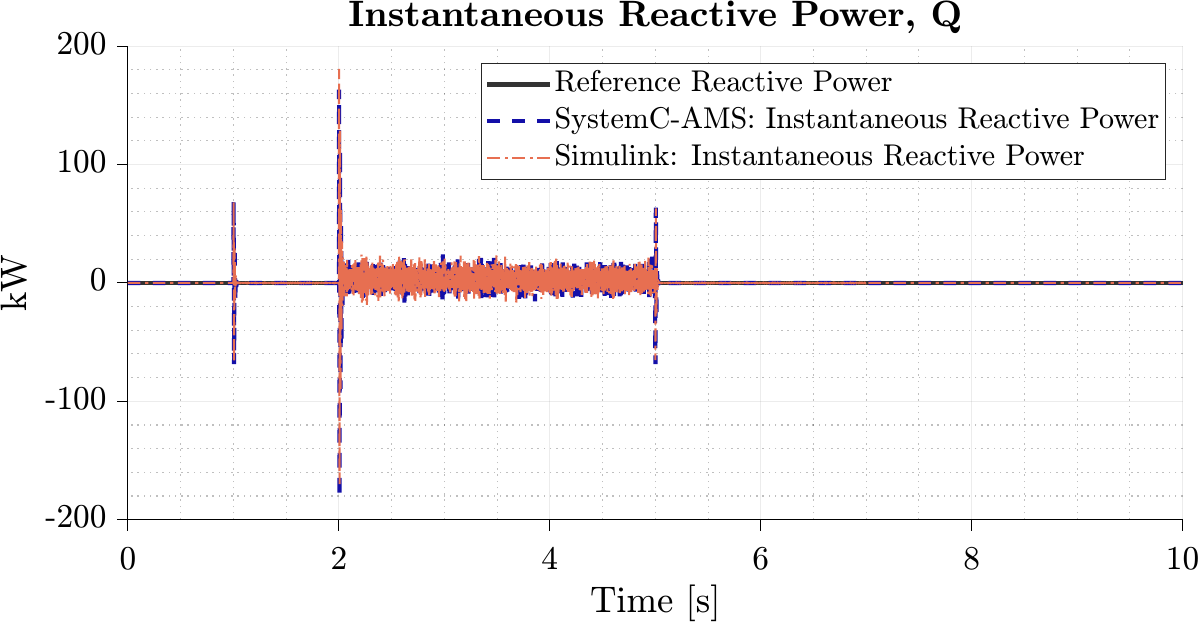}
\caption{}
\end{subfigure}
\begin{subfigure}{0.8\linewidth}
\centering
\includegraphics[width=1.0\linewidth, trim={0.0cm 0.0cm 0.0cm 0.0cm},clip]{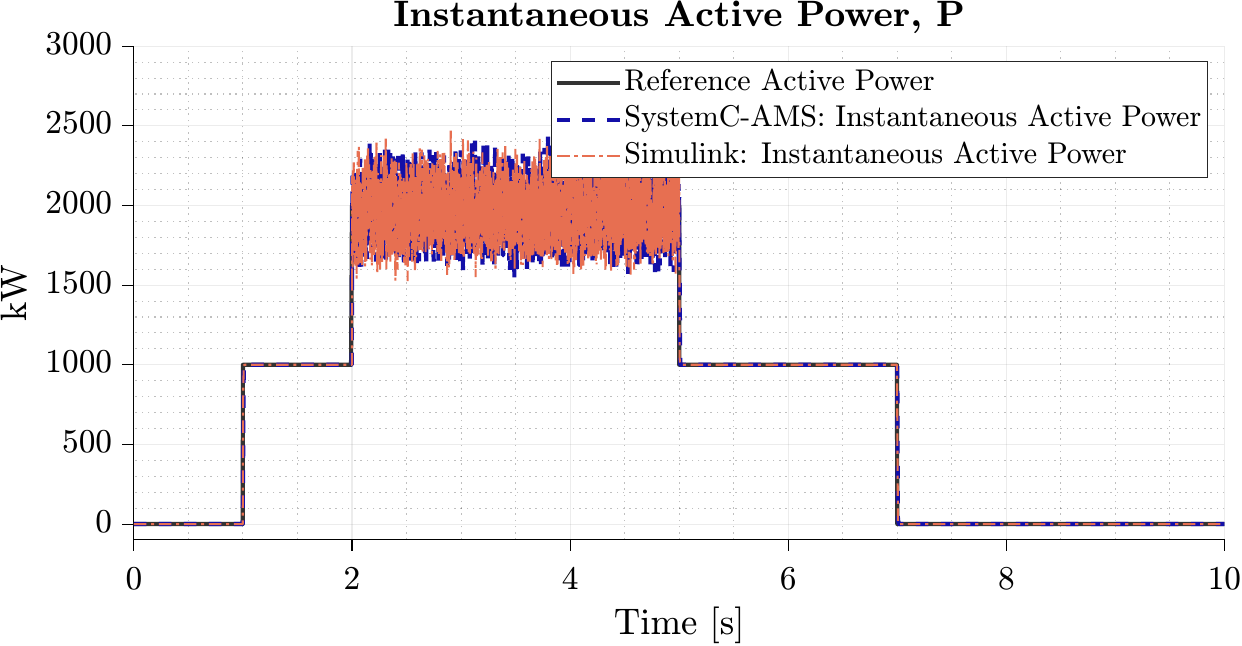}
\caption{}
\end{subfigure}
\caption{\textbf{\textit{Simplified GFL inverter feedforward model: step-response.}} \textbf{Top: }Instantaneous reactive power plot comparison in SystemC-AMS simulation vs. Simulink Simulation. We also plot the reference reactive power. \textbf{Bottom: }Instantaneous active power plot in SystemC-AMS simulation vs. Simulink Simulation. The power tracking capability of the GFL inverter feedforward model is poorer than that with inner current loops as we can see in Figure~\ref{fig:power_plot_lpf} that uses a low-pass filter to add inertial response from the inverter.}
\label{fig:power_plot_feedfwd}
\end{figure}

In the case of the simplified GFL inverter without inner current loops, the RMS error of the instantaneous active power compared to the reference active power for SystemC-AMS simulation is $85.78328~\textrm{kW}$, and $85.87311~\textrm{kW}$ for Simulink simulation which is roughly within $1\%$ the initial step amplitude of $1000~\textrm{kW}$. The reactive power RMS error was found to be $4.35805~\textrm{kVar}$, and $4.36262~\textrm{kVar}$ for SystemC-AMS simulation and Simulink simulation respectively. In addition, we also assess the RMS value for reactive power between two simulation methods around the time when the first transient occurs. The first transient takes approximately 0.2 seconds to stabilize. RMS error between Simulink and SystemC-AMS simulink, while transient lasted, came out to be $0.04996~\textrm{kVar}$.

The simplified GFL inverter feedforward model provides poor tracking of the reference active and reactive power, however, it is easier to implement. With the feedforward model, the instantaneous active power compared to the reference active power for SystemC-AMS
simulation, and Simulink simulation are $80.39742~\textrm{kW}$, and 80.20730 kW for Simulink
simulation. The reactive power RMS error are $5.92436~\textrm{kVar}$, and $5.8789~\textrm{kVar}$ for SystemC-AMS simulation and Simulink simulation respectively. The result is summarized in Table~\ref{tab:step}.

\subsection{Ramp-function Reference Power}
\label{sec:ramp_fn}
Further, we study the ramp response of our proposed GFL inverter design by modifying our step signal such that reference input changes uniformly rather than abruptly. GFL inverter without inner
current loops provides an RMS error of $23.5076~\textrm{kW}$ for active power when compared to the reference active power, and $1.1764~\textrm{kVar}$ for reactive power when compared to the reactive power. We do not observe transients when our reference input changes uniformly. The simplified GFL inverter feedforward model provides poorer tracking compared to the one without inner current loops similar to the one observed with step-response. We obtained the RMS error of $4.8629~\textrm{kVar}$ for reactive power and $80.1555~\textrm{kW}$  for active power when compared to their respective reference inputs. Reactive power and active power plot from the simulation study is provided in Figure

\begin{figure}[htpb]
\captionsetup[subfigure]{aboveskip=0pt,belowskip=1pt}
\centering
\begin{subfigure}{0.8\linewidth}
\centering
\includegraphics[width=1.0\linewidth, trim={0.0cm 0.0cm 0.0cm 0.0cm},clip]{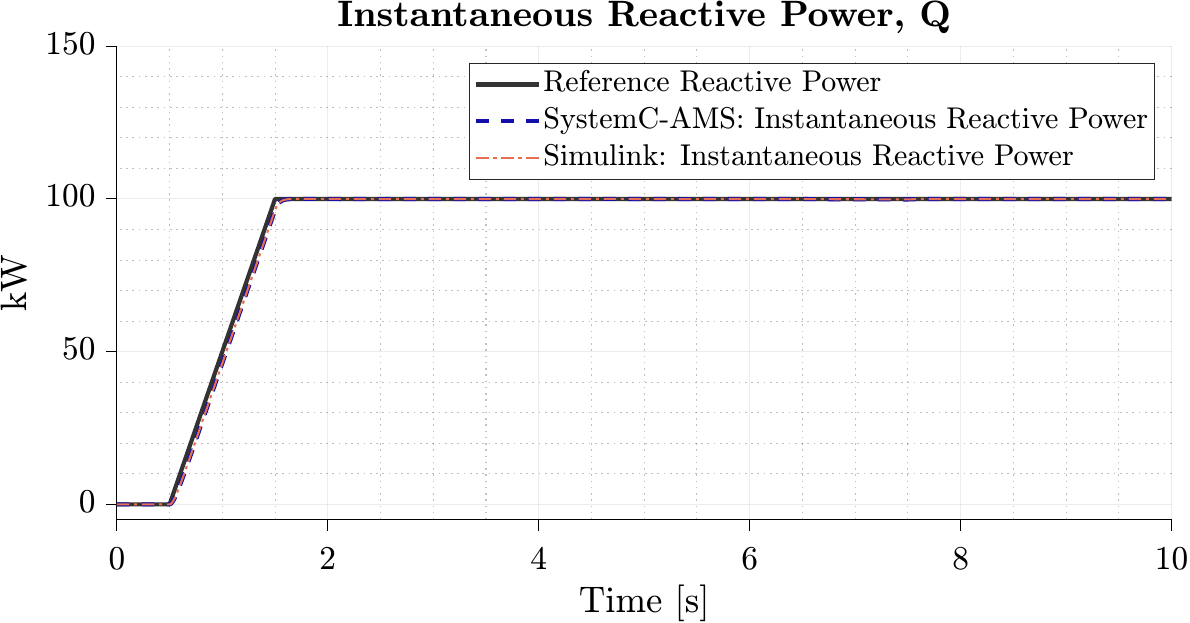}
\caption{}
\end{subfigure}
\begin{subfigure}{0.8\linewidth}
\centering
\includegraphics[width=1.0\linewidth, trim={0.0cm 0.0cm 0.0cm 0.0cm},clip]{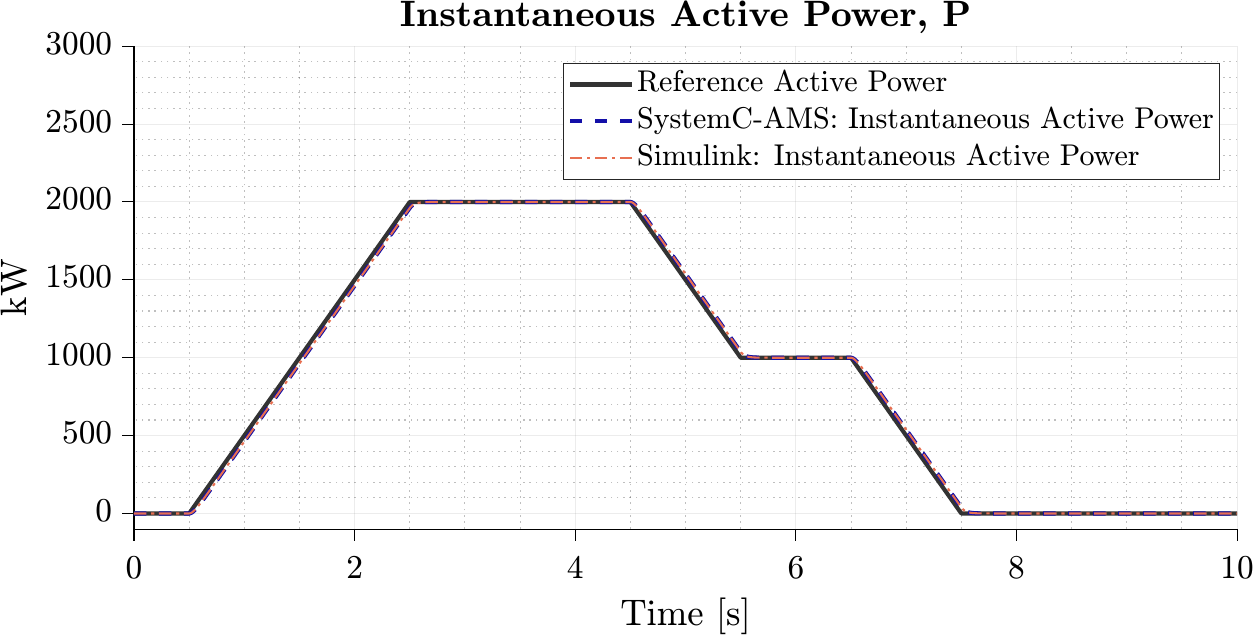}
\caption{}
\end{subfigure}
\caption{\textbf{\textit{Simplified GFL inverter without inner current loops: ramp response.}} \textbf{Top:} reactive power comparison for Simulink and SystemC-AMS.  \textbf{Bottom: } active power comparison for Simulink and SystemC-AMS. As the power level gradually changes, we do not observe EMT. However, both simulation method yields approximately the same results.}
\label{fig:ramp_power_plot_lpf}
\end{figure}
\begin{figure}[htpb]
\captionsetup[subfigure]{aboveskip=0pt,belowskip=1pt}
\centering
\begin{subfigure}{0.8\linewidth}
\centering
\includegraphics[width=1.0\linewidth, trim={0.0cm 0.0cm 0.0cm 0.0cm},clip]{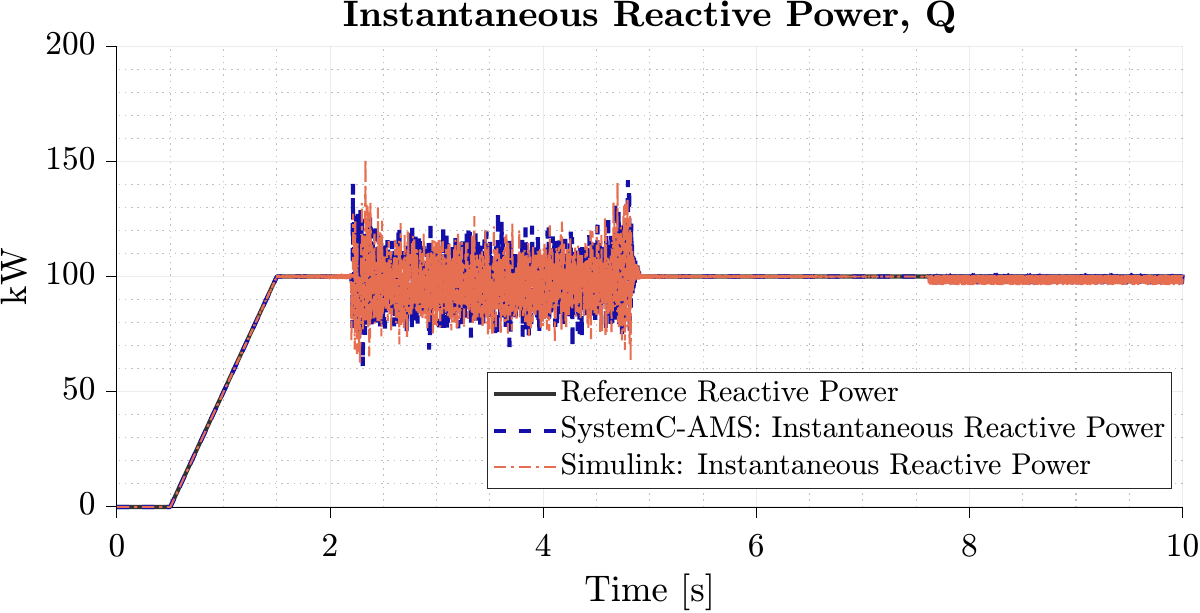}
\caption{}
\end{subfigure}
\begin{subfigure}{0.8\linewidth}
\centering
\includegraphics[width=1.0\linewidth, trim={0.0cm 0.0cm 0.0cm 0.0cm},clip]{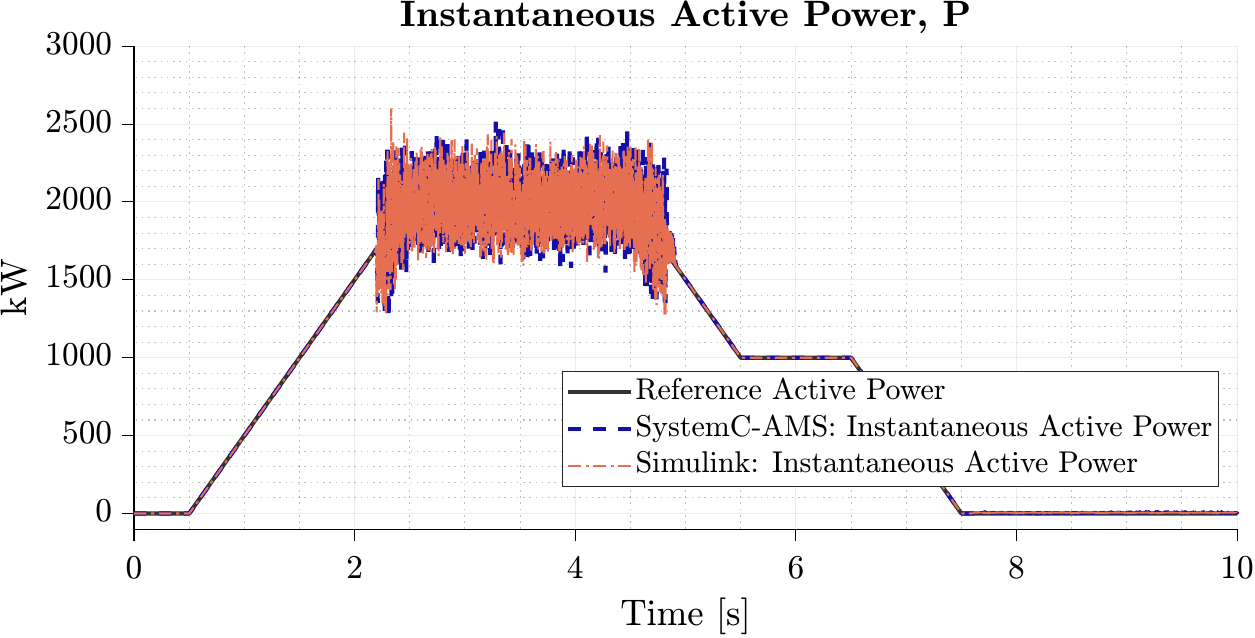}
\caption{}
\end{subfigure}
\caption{\textbf{\textit{Simplified GFL inverter feedforward model.}} \textbf{Top:} reactive power comparison for Simulink and SystemC-AMS.  \textbf{Bottom: } active power comparison for Simulink and SystemC-AMS. Without the presence of a low pass filter as seen in the GFL inverter without inner current loop design, power tracking is poorer and is observed in both method simulation tools.}
\label{fig:ramp_power_plot_feedfwd}
\end{figure}

From the comparative result, we find out that in the case of both Simulink as well as SystemC-AMS, the instantaneous power was within $1\%$ of the reference value. Hence, we conclude that SystemC-AMS is capable of providing a high-fidelity simulation tool for simulating microgrid design.

\subsection{Run-time Performance}
\begin{figure}[htpb]
\centering
\includegraphics[width=1.0\linewidth, trim={0cm 0.0cm 0cm 0.0cm},clip]{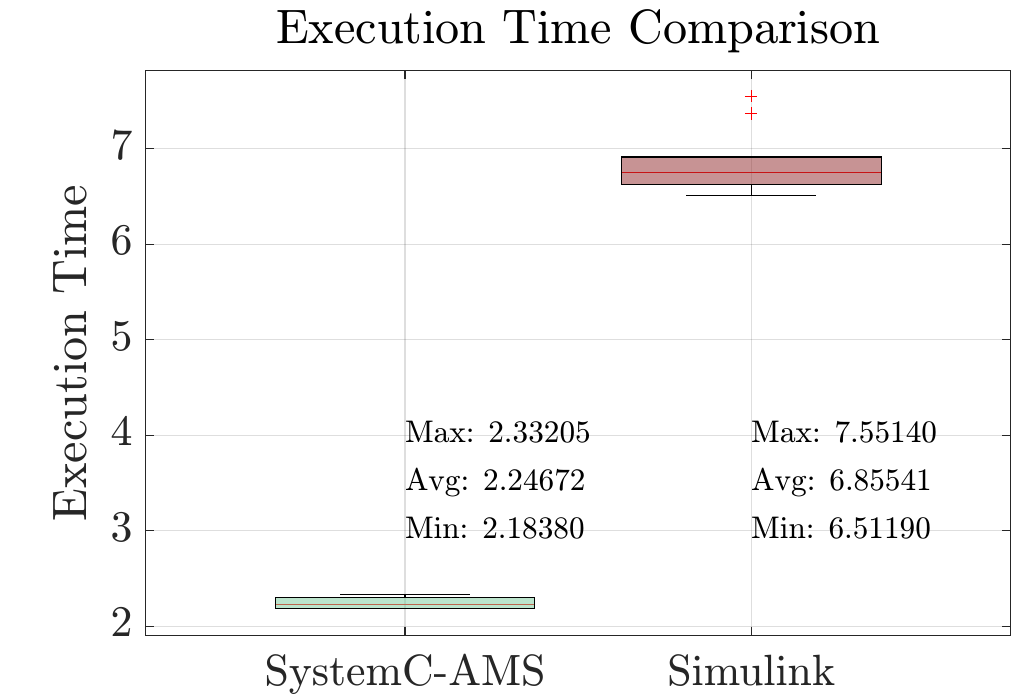}
\caption{Execution time comparison of simulation conducted in SystemC-AMS and Simulink We observe that the simulation conducted in SystemC-AMS is approximately 3x faster than one done in Simulink.}
\label{fig:boxplot}
\end{figure}

The primary objective of conducting simulation in SystemC-AMS is to perform faster simulation than the state-of-the-art simulation done in Simulink. Once we established the correctness of the simulation results as discussed in Section~\ref{sec:step_fn} and Section~\ref{sec:ramp_fn}, we calculated how much wall-clock time elapsed while running the simulation with a duration of $10~\textrm{sec}$. We repeated the experiment 10 times and recorded the distribution of wall-clock time elapsed for simulation done using Simulink as well as SystemC-AMS.
Our calculation shows that the simulation done in SystemC-AMS was approximately three times faster than the one done in Simulink. Figure~\ref{fig:boxplot} illustrates the execution performance using a boxplot diagram.

\section{Real-time Simulation of a DC Microgrid Model with Constant Resistive Load}
\label{sec:real_time}
\begin{figure}[htpb]
\centering
\includegraphics[width=1.0\linewidth, trim={0cm 0.0cm 0cm 0.0cm},clip]{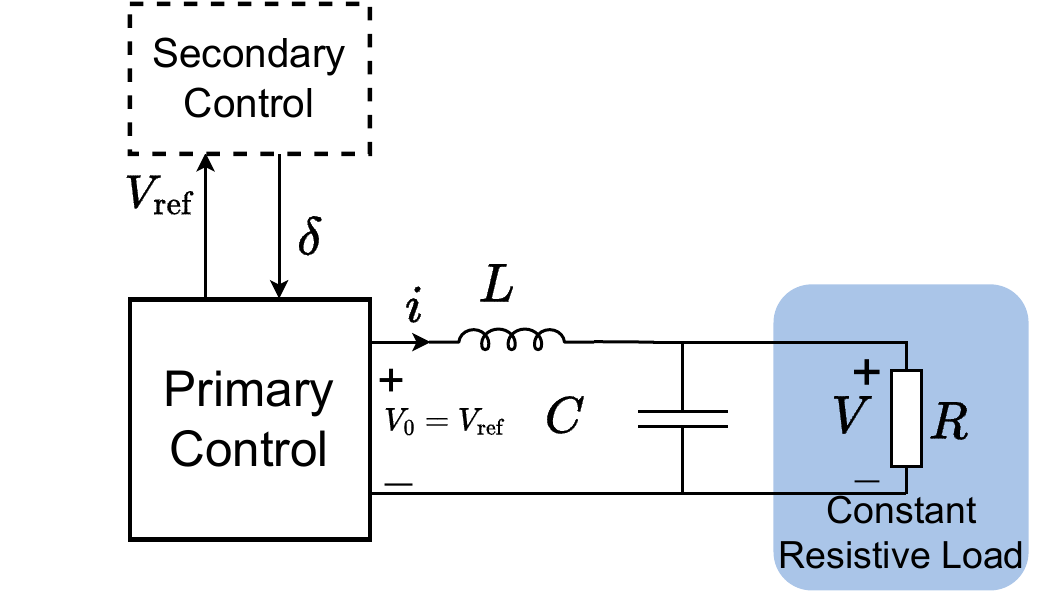}
\caption{The model of a DC microgrid with a steady resistive load is considered. The primary controller, known as the droop controller, sets the reference voltage using a nominal voltage and the output current. If needed, a secondary controller can be employed to fine-tune the nominal voltage set by the primary controller in order to control the voltage of the microgrid, with an extra correction term $\delta(t)$.}
\label{fig:constant_resistive_load}
\end{figure}

In another case study, we decouple the controller from the microgrid plant to facilitate a real-time simulation. We use a DC microgrid with a primary control and a secondary control where the DC microgrid along with the primary control acts as a plant while the secondary control is a separate model communicating with the plant through ZeroMQ communication API~\cite{hintjens2013zeromq}. We model a single-bus DC microgrid consisting of a DC-DC converter, an inductor, a capacitor, and a load~\cite{tu2023impact} (see Figure \ref{fig:constant_resistive_load}). The inductor and capacitor together are equivalent to the line impedance, filters, and DC bus capacitor.
The equation describing the DC microgrid is 
 \spliteq
 {
 \label{eq:dcmicrogrid}
 V_{\text{ref}}(t) - V(t) & = L \cfrac{di}{dt}\\
i(t)-\cfrac{V(t)}{R} & = C \cfrac{dV}{dt}
 }
 where $i$ is the inductor current;  $V$ is the capacitor voltage; $V_0$ and $V_{\text{ref}}$ are the converter's output voltage and reference voltage, respectively. We also assume that the output voltage can track the reference voltage as accurately as possible due to the converter's inner current loop which is $V_0 = V_\textrm{ref}$. $V_\textrm{ref}$ is generated by a primary controller (a droop controller) as follows
 \spliteq
 {
 \label{eq:drropcontrol}
V_{\text{ref}}(t) = V_n- k\cdot i(t)
}
where $k$ is the droop gain, and $V_n$ is the nominal voltage. Any fluctuation in the current $i(t)$, due to the constant resistive load, directly affects the reference voltage $V_{\text{ref}}$.  In order to keep the voltage of a droop-controlled DC microgrid at its nominal level, a secondary controller can be implemented. This controller adjusts the nominal voltage $V_n$ using a correction term $\delta(t)$. In the context of this article’s use case, both the primary and secondary controllers are regulated by the subsequent equation:
\begin{equation}
\begin{split}
\label{eq:secondary_controller}
V_n'(t) & = \delta(t) + V_n\\
V_{\text{ref}}(t) & = V_n'(t)- k\cdot i(t)\\
\delta(t) & = \int k_s(V_n - V_{\text{ref}}(t)) dt
\end{split}
\end{equation}
where $k_s$ is the secondary control gain. Secondary controllers typically operate at a frequency ranging from 1 to 100 Hz~\cite{Tu2020}, a rate that is compatible with real-time execution using SystemC-AMS. We set the nominal voltage of $V_n = 200~\textrm{V}$, $k=0.2$, and $k_s = 0.75$.

\begin{figure}[htpb]
\centering
\includegraphics[width=1.0\linewidth, trim={0cm 0.0cm 0cm 0.0cm},clip]{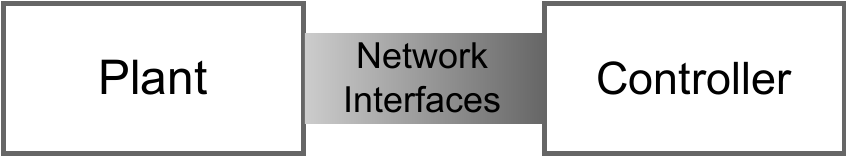}
\caption{A microgrid simulation with the plant and controller exchanging data through a communication interface. In the context of this paper, we use ZeroMQ for communication.}
\label{fig:communication_interface}
\end{figure}

\begin{figure}[htpb]
\centering
\includegraphics[width=1.0\linewidth, trim={0cm 0.0cm 0cm 0.0cm},clip]{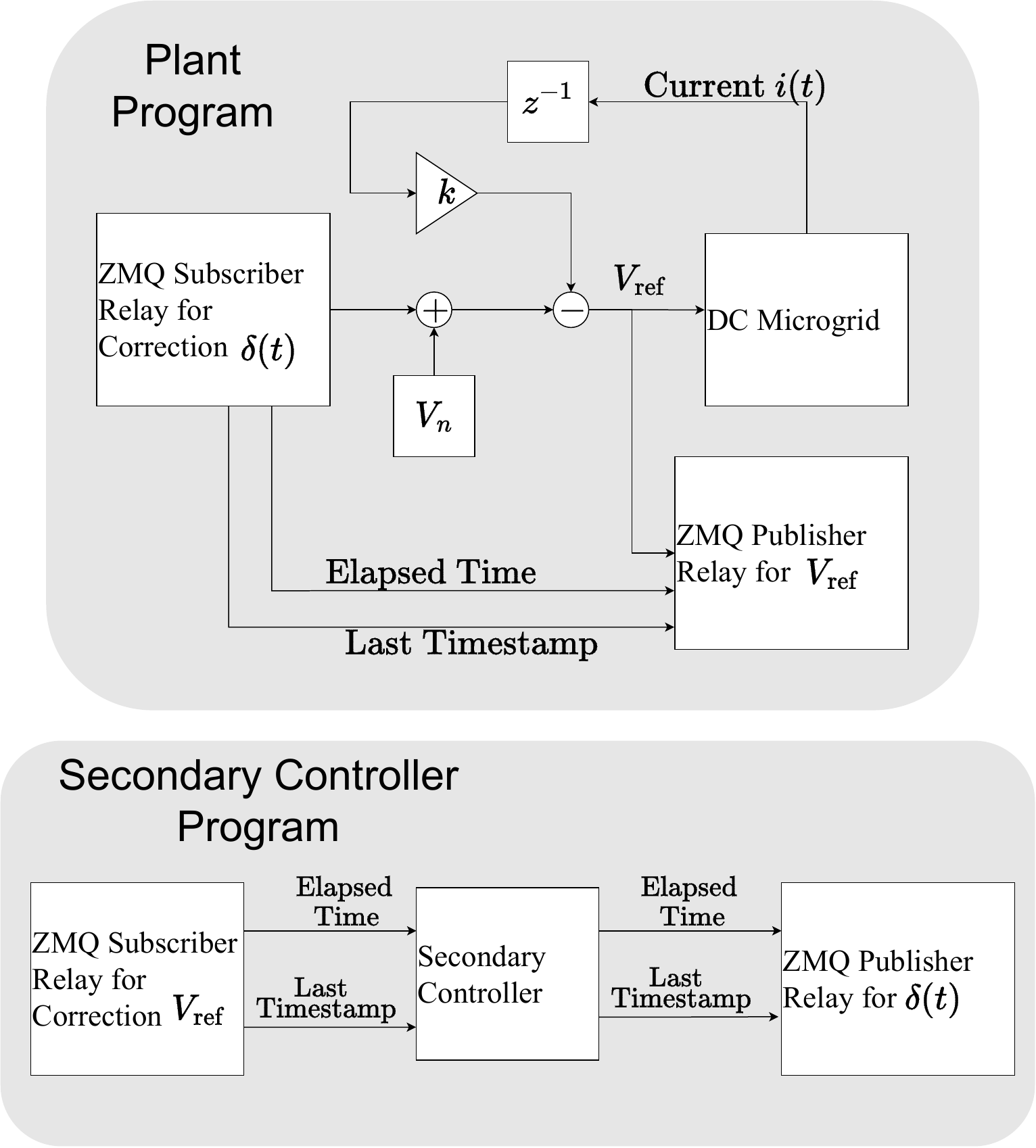}
\caption{A diagram of the DC microgrid features a primary controller, along with ZeroMQ subscriber and publisher relay modules. The subscriber relay module retrieves data from the secondary controller, logging the timestamp of the data retrieval. This information is then passed to the publisher relay module. Using the most recent timestamp and the elapsed time, the publisher relay module calculates the necessary wait time to simulate real-time behavior. Additionally, an algebraic loop is prevented by incorporating a delay unit.}
\label{fig:DCMicrogrid}
\end{figure}

A schematic of our approach for real-time simulation is shown in Figure~\ref{fig:communication_interface}.
As SystemC-AMS is designed for fast simulation, only limited by the device's capability, we create a customized TDF module called the relay module that uses ZeroMQ API to schedule the message exchange between the plant and the controller using ZeroMQ subscribers and publishers. The module corresponding to the controller and the plant measures the computation time to produce the new result that is propagated by the relay modules. The relay modules deliver the new result to the other party only if time corresponds to the step size of the simulation that has elapsed since the start of the new calculation. Note that in this case, real-time simulation is only possible if the time required to perform a calculation by a SystemC-AMS module is less than the specified step size of the simulation. In addition to using ZeroMQ, we use \texttt{PREEMPT\_RT} kernel patch in the Ubuntu and execute the binaries corresponding to the simulation with the highest scheduling priority. A schematic of the implementation is illustrated in Figure~\ref{fig:DCMicrogrid}. Please refer to a paper~\cite{bhadani2023wsc} for additional details on this topic.

\begin{figure}[htpb]
\centering
    \includegraphics[angle=0,origin=c,trim={0cm 0.0cm 0.0cm 0.0cm},clip,width=1.0\linewidth]{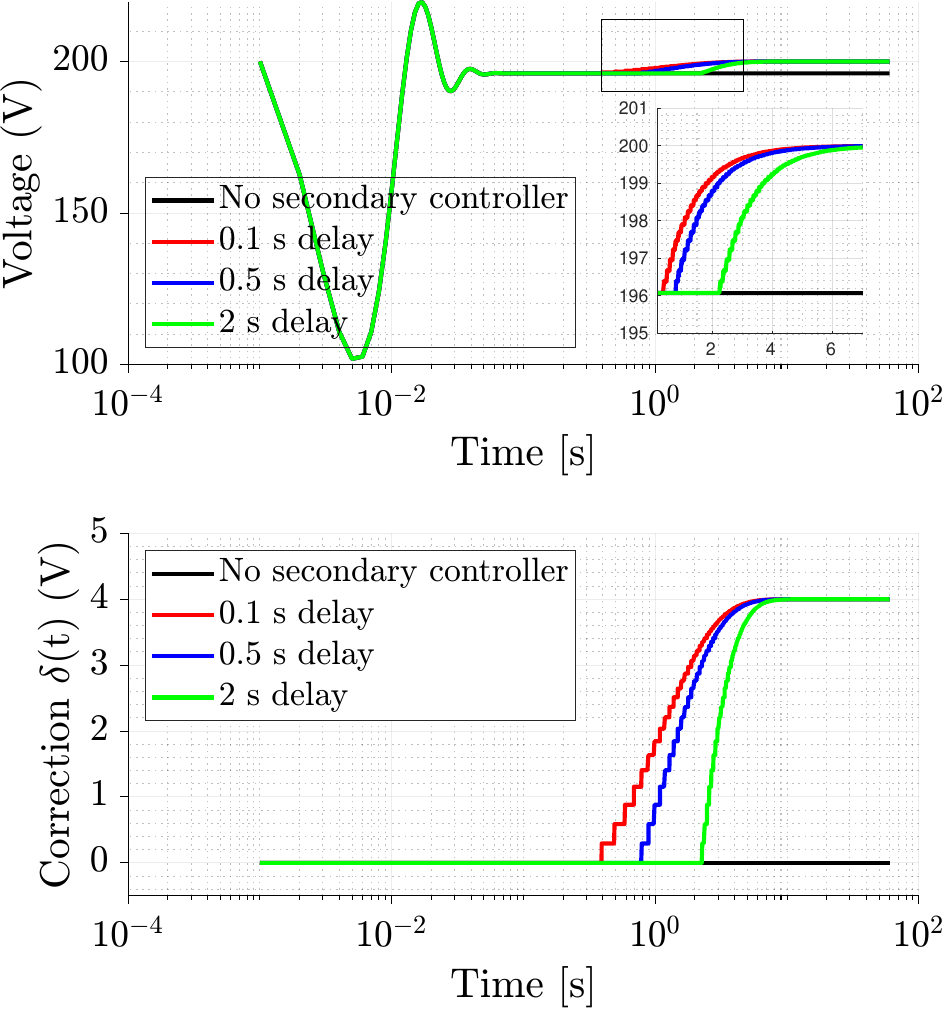}
\caption{Simulation results of a DC microgrid, where the primary controller acts as a plant in a SystemC-AMS simulation, are presented. These results include voltage traces and the correction term $\delta(t)$. The output voltage $V_0$ which is measured as the adjusted reference voltage from Equation~\eqref{eq:secondary_controller} starts at $200~V$ (shown in the top figure), but it falls due to fluctuation in the current caused by the resistive load. The secondary controller compensates for the decrease in the output and over time voltage stabilizes to the set nominal voltage. Different scenarios of secondary controller execution are demonstrated, starting with a scenario where it is completely absent. For each case of $0.1$ s, $0.5$ s, and $2.0$ s delay, we zoom in on a segment between $0.2$~s to $7.0$~s of the voltage plot in the inset on the top subplot (marked with a box on the full plot). We observe that the voltage stabilizes later if the secondary controller comes online with some delay. Note that the x-axis is log-scaled.}
\label{fig:Delayed_DCGrid_Plot}
\end{figure}

The plant simulation was carried out with a time step of 1 ms, whereas the secondary controller operated with a time step of 100 ms. These values are representative of how the primary controller (a part of the plant) and the secondary controller operate in the real world. Looking at Figure~\ref{fig:Delayed_DCGrid_Plot}, 
the output voltage $V_0$ is measured as the adjusted reference voltage from Equation~\eqref{eq:secondary_controller} and starts at $200~V$ (shown in the top subplot of Figure~\ref{fig:Delayed_DCGrid_Plot}). However, it fails to maintain the nominal value due to fluctuation in the current caused by the resistive load. The secondary controller compensates for the decrease in the output and over time voltage stabilizes to the set nominal voltage. During its operation, the plant keeps the old value of the signal until receives a new value from the secondary controller. Such real-time simulation is also useful when the secondary controller is implemented by some other program or in fact it can be a hardware implementation to allow hardware-in-the-loop simulation. Initially, the plant may execute in an open-loop manner and once the controller is online, it is expected to stabilize the plant if designed correctly. To study such behavior, we induce the delayed start of the controller. In the absence of the controller, the reference voltage doesn't reach the specified nominal voltage of $200~\textrm{V}$. When the secondary controller is executed with a delayed start, the settling time is achieved at a later time. When the start is delayed by $0.1$~s, the reference voltage’s settling time is $0.986$ s. If the start is delayed by $0.5$ s, the settling time becomes $1.383$ s, and with a delay of $2.0$ s in the start, the settling time is observed to be $2.869$ s. The result from this simulation is plotted in Figure~\ref{fig:Delayed_DCGrid_Plot}.

\begin{figure}[htpb]
\centering
\includegraphics[width=1.0\linewidth, trim={0.0cm 0.0cm 0.0cm 0.0cm},clip]{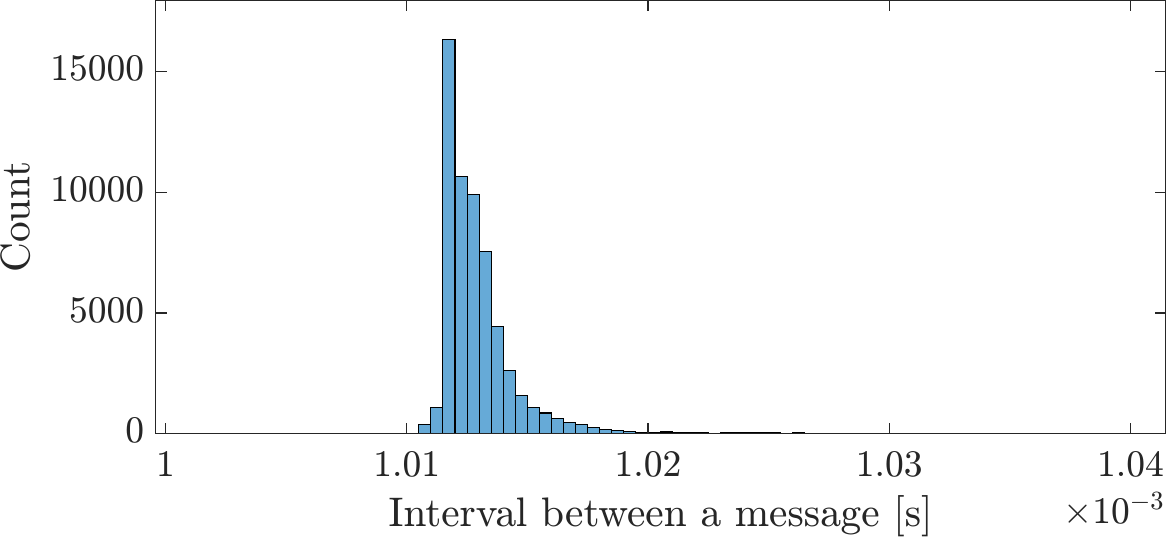}
\caption{Histogram showing intervals between the timestamps of message publications.}
\label{fig:histogram_pub_time}
\end{figure}

To assess the efficacy of real-time simulation, we also logged the time-stamp of the reference voltage message published by the ZeroMQ relay module.  The DC microgrid plant simulation was configured with a time-step of $1$ ms. In a perfect real-time system, we would anticipate the average time difference to be $1$ ms with no standard deviation. However, the simulation data revealed that the median time difference was $1.01259$ ms, the average was $1.01728$ ms, and there was a standard deviation of $0.08753$ ms. A Histogram of the message interval is provided in Figure~\ref{fig:histogram_pub_time}.

\section{Conclusion and Future Outlook}
In this paper, we have presented our work on the model-based design of microgrid components using SystemC-AMS, constructing a DC microgrid, and a microgrid design using GFL inverters. We conducted a simulation study of GFL with an ideal main grid. Our approach demonstrated that SystemC-AMS can perform a fast simulation, exhibits EMT phenomenon, and can interface with external libraries. Additionally, we introduced a real-time simulation method that incorporates a communication component, using the ZeroMQ C++ library for message exchange between the plant and controller simulations. This strategy allows SystemC-AMS to function as a digital twin for microgrids, facilitating hardware-in-the-loop experiments with hardware prototypes to refine control algorithms. Our future work involves expanding grid components in SystemC-AMS to study microgrids at scale and demonstrating the capabilities of real-time simulation in conjunction with hardware components to regulate grid signals under various conditions. In the follow-up of the current work, we will test the SystemC-AMS implementation of a microgrid with middleware control applications implemented in RIAPS~\cite{ghosh2023distributed, eisele2017riaps}.

\section*{Acknowledgement}
The information, data, or work presented herein was partly funded by the Advanced Research Projects Agency-Energy (ARPA-E), U.S. Department of Energy, under Award Number DE-AR0001580. The views and opinions of authors expressed herein do not necessarily state or reflect those of the United States Government or any agency thereof.

\bibliographystyle{IEEEtran}
\bibliography{references}
\end{document}